\documentclass[12pt,letterpaper]{article}
\usepackage[a4paper, total={7in, 10in}]{geometry}

\usepackage{graphicx}
\usepackage{helvet}
\usepackage{authblk}
\usepackage{hyperref}
\usepackage{amsmath} 
\usepackage{amssymb} 
\usepackage{orcidlink} 
\usepackage[super,comma,sort&compress]  
   {natbib}
\usepackage[right]{lineno} \linenumbers

\usepackage{graphicx}%
\usepackage{multirow}%
\usepackage{amsmath,amssymb,amsfonts}%
\usepackage{amsthm}%
\usepackage{mathrsfs}%
\usepackage[title]{appendix}%
\usepackage{xcolor}%
\usepackage{textcomp}%
\usepackage{manyfoot}%
\usepackage{booktabs}%
\usepackage{algorithm}%
\usepackage{algorithmicx}%
\usepackage{algpseudocode}%
\usepackage{listings}%
\usepackage{siunitx}%
\usepackage{makecell}
\usepackage{booktabs}  
\usepackage{float}     
\usepackage{arydshln}

\makeatletter
\renewcommand{\maketitle}{\bgroup\setlength{\parindent}{0pt}
\begin{flushleft}
  \textbf{\@title}
  
  \@author
\end{flushleft}\egroup}
\makeatother

\title{Entanglement distribution over 155~km metropolitan fiber using a CMOS-compatible silicon chip}

\author[1,*]{Jinyi Du}
\author[1,**]{Xingjian Zhang}
\author[2]{George F.R. Chen}
\author[2]{Hongwei Gao}
\author[2,3,***]{Dawn T. H. Tan}
\author[1,4,****,$\dagger$]{Alexander Ling}

\affil[1]{Centre for Quantum Technologies, National University of Singapore, 3 Science Drive 2, Singapore 117543, Singapore}
\affil[2]{Photonics Devices and System Group, Singapore University of Technology and Design, 8 Somapah Road, Singapore 487372, Singapore}
\affil[3]{Institute of Microelectronics, Agency for Science, Technology and Research (A*STAR), 2 Fusionopolis Way, \#08-02 Innovis Tower, Singapore 138634, Singapore}
\affil[4]{Department of Physics, National University of Singapore, 2 Science Drive 3, Singapore 117551, Singapore}

\affil[*]{jinyidu@u.nus.edu}
\affil[**]{xingjianzhang@u.nus.edu}

\affil[***]{dawn\_tan@sutd.edu.sg}
\affil[****]{alexander.ling@nus.edu.sg}
\affil[$\dagger$]{Lead contact}

\begin{document}

\maketitle

\section*{SUMMARY}

Transmitting entangled states over long distances is crucial for developing quantum networks. Previous demonstrations using satellites or fibers relied on photon pairs generated from bulk crystal arrangements. Polarization entanglement distribution based on CMOS-compatible silicon chips has long been restricted to lab-scale demonstrations spanning only a few meters, due to the difficulty of achieving sufficient off-chip brightness. We report a silicon chip platform that provides an off-chip entangled photon pair brightness ranging from 8,000 to 460,000 pairs per second, exceeding previous reports by three orders of magnitude. The entanglement fidelity reaches 99.85(6)\% and 97.90(3)\%, respectively. After addressing key challenges in long-distance entanglement distribution over deployed fiber, including phase drift and chromatic dispersion, entangled photons were successfully distributed over 155~km (66~dB loss). These results demonstrate that CMOS-compatible silicon chips can perform competitively with bulk crystal sources and represent an important step toward scalable, chip-based quantum networks.
\section*{KEYWORDS}
Entanglement distribution, CMOS-compatible silicon chip, Quantum network

\section*{INTRODUCTION}

Entanglement distribution will be important for building quantum networks that provide secure communications or novel services such as blind quantum computing. Impressive distances have been achieved using satellite-based entanglement technology \cite{liao2017satellite,yin2017satellite,villar2020entanglement,yin2020entanglement}, or with terrestrial free-space links using telescopes \cite{ursin2007entanglement,yin2012quantum}.  For dense urban environments, however, optical fibers provide an alternative channel that is not limited by atmospheric conditions. 

A common strategy for fiber-based transmission is to use time-bin entanglement \cite{marcikic2004distribution,takesue2005generation,honjo2007long,dynes2009efficient,inagaki2013entanglement}, which is robust against polarization drift in aerial fibers \cite{shi2023towards}. However, time-bin entanglement typically relies on unbalanced interferometers which demands precise phase stabilization, and structurally leads to a 3~dB measurement loss.

In the stable underground fiber networks, polarization entanglement offers a more direct and flexible platform for state preparation and analysis. It is also inherently compatible with satellite-to-ground free-space links  and supports hybrid architectures that span fiber and free-space channels \cite{jin2010experimental,yin2012quantum,ursin2006free,ma2012quantum}. The primary challenge for polarization-entangled photons is polarization mode dispersion (PMD) in fiber, but effective mitigation strategies have recently been proposed \cite{xingjian2025polarization,rodimin2024impact}. 

Bulk nonlinear crystal-based sources such as PPLN \cite{wengerowsky2019entanglement,wengerowsky2020passively,neumann2022continuous} and PPKTP \cite{rahmouni2024metropolitan} have successfully generated bright polarization-entangled photons, which have been distributed over distances up to 248~km using deployed fiber \cite{wengerowsky2019entanglement,wengerowsky2020passively,neumann2022continuous} and 404~km using fiber spools \cite{zhuang2024ultrabright}. In contrast, polarization-entangled photons from CMOS-compatible silicon nanophotonic chips have long been limited to lab-scale demonstrations over just a few meters \cite{llewellyn2020chip,zheng2023multichip}. To overcome the link loss over long fiber stretches, the photon pair sources need to be very bright and to-date silicon photonic chips have not provided high off-chip brightness due to coupling losses \cite{du2024demonstration,suo2015generation, takesue2008generation, li2017chip, sharma2022silicon}, whereas bulk optical systems are several orders of magnitude brighter than their chip-based counterparts. 

In this work, we present a system-level optimization approach that addresses several challenges. This enables silicon photonic chips to operate in a regime previously dominated by bulk-crystal systems. 
We report a bright polarization entangled photon pair source based on a silicon nanophotonic chip that can distribute entanglement through approximately 66 dB of loss, corresponds to 155 km of deployed optical fiber in a metropolitan network. By using a novel edge coupler, the silicon chip is placed in a Sagnac configuration and enables direct observation of entangled photon pairs at a rate of between 8,000 to 460,000 pairs per second. This off-chip brightness is three orders of magnitude higher prior silicon-based reports \cite{li2017chip,takesue2008generation,suo2015generation,miloshevsky2024cmos,llewellyn2020chip,zheng2023multichip,wen2023polarization,zhang2019generation}. The entanglement fidelity is 99.85(6)\% and 97.90(3)\% respectively.

Measuring one photon locally, and transmitting the other over 93 km of deployed fiber (link loss of 40 dB), achieves a count rate of 132 pairs per second with an entanglement fidelity of 93.3(3)\%. To mitigate chromatic dispersion over long fiber distances, we implement nonlocal dispersion compensation \cite{neumann2022continuous,chua2022fine,neumann2021experimentally}, improving coincidence-to-accidental ratio (CAR) and fidelity upper limit by 32\%. 
The source can be pumped harder to enable transmission of entangled photons over 155 km of deployed fiber (link loss of 66 dB) at a rate of 0.7 pairs per second, with an entanglement fidelity of 87.6(5)\%.
This result demonstrates that the CMOS chip sources, can compete with crystal-based systems for long-range quantum networking.

\section*{RESULTS}

\section*{On-chip entangled photon generation}\label{subsec1}
\begin{figure}[H]
	\centering
	\includegraphics[width=1\textwidth]{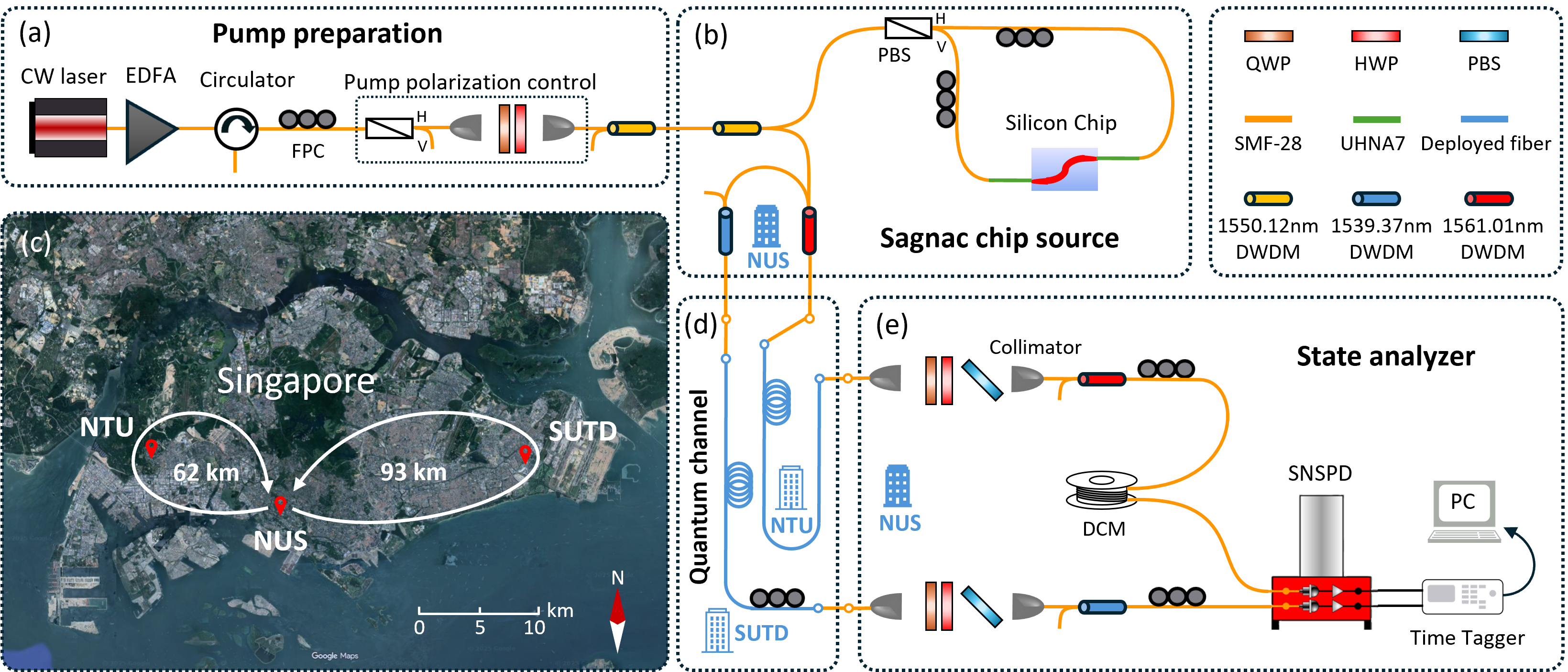}
	\caption{\textbf{Entanglement distribution over 155 km deployed fiber.} (a) A continuous-wave laser at 1550.12 nm is amplified by an erbium-doped fiber amplifier (EDFA), with a circulator used to isolate the back-propagating beam from the Sagnac interferometer. The polarization control unit allows fine-tuning of the pump splitting ratio within the Sagnac loop. (b) The amplified laser generates polarization-entangled photon pairs via spontaneous four-wave mixing (SFWM), enabled by the Sagnac configuration. The nondegenerate photon pairs are wavelength-separated using a dense wavelength-division multiplexer (DWDM). (c) One fiber runs from the National University of Singapore (NUS) to the Singapore University of Technology and Design (SUTD) and back to NUS via a second fiber follows a circuitous path, resulting in a total length that exceeds the straight-line distance between these locations. Another fiber runs from NUS to Nanyang Technological University (NTU) and back to NUS via a second fiber. Satellite imagery: Google Maps (map data ©2025).
		(d) The signal photon (1539.37 nm) travels through the SUTD route while the idler photon (1561.01 nm) travels through the NTU route. The deployed fibers can be replaced with patchcord fiber when measuring the source entanglement fidelity. (e) The photons are eventually detected by a superconducting nanowire single-photon detector (SNSPD) at NUS. To correct for fiber birefringence, fiber polarization controllers (FPC) and quarter-wave plates (QWP) are utilized. Half-wave plates (HWP) facilitate the switching between horizontal (H), vertical (V), diagonal (D), and anti-diagonal (A) polarizations. Polarization beam splitters (PBS) in the setup are used to project the state onto the desired polarization for analysis. A dispersion compensation module (DCM) is placed on the idler photon arm before the SNSPD to provide nonlocal dispersion compensation.}\label{setup}
\end{figure}

The experimental setup for generating polarization entangled photons is illustrated in Fig.~\ref{setup}.(a, b). The test utilizes a nanophotonic chip that is an 8 mm long silicon waveguide, equipped with edge couplers at both ends, each exhibiting a loss of 0.64 dB. 
The silicon chip is integrated into a Sagnac fiber loop consisting of a PBS and two fiber polarization controllers (FPC) on each arm. A continuous-wave (CW) laser with a wavelength of 1550.12 nm first passes through a standard dense wavelength division multiplexer (DWDM) filter and is then evenly split by the PBS. The vertically polarized (V) beam is converted to horizontal polarization (H) with FPC before entering the chip. A short section of \SI{2.4}{\micro\meter} small-core fiber (UHNA7), less than 1 cm in length, is spliced to the tip of coupling fiber to assist mode conversion with a splicing loss of 0.03 dB. 

A pump beam injected into the chip causes spontaneous four wave mixing (SFWM) to occur. A pair of pump photons at frequency $\omega_{pump}$ are annihilated, generating a new photon pair, following the energy conservation equation $2\omega_{pump} = \omega_{signal} + \omega_{idler}$ \cite{boyd2008nonlinear}. To create entangled photon pairs, the pump is injected from both ends of the waveguide simultaenously, leading to the generation of photon pairs propagating in both directions. These pairs can be collected from either side of the waveguide using the same fiber as pump laser. One pair is horizontally (H) polarized and the other pair is converted to vertically (V) polarized by FPC. 
As the H and V state photons travel through the same fiber, no phase shift occurs between the two arms.  By superposing the H and V polarized photon pairs from both paths at the PBS, the maximally entangled Bell state is constructed at the output port of the PBS:
\begin{equation}
	|\Phi\rangle = \frac{1}{\sqrt{2}} (|HH\rangle + |VV\rangle),
\end{equation}
The entangled photons are reflected by a DWDM module centered at 1550.12 nm to the reflection arm. A narrow band portion (0.8 nm) of photon pairs are then separated using 100 Ghz DWDM modules centered at 1561.01 nm (idler) and 1539.37 nm (signal). The broadband nature of the entangled photons generated from the SFWM process in the straight Si waveguide allows for the cascading of additional filters to multiplex more channels of entangled photons \cite{li2017chip,appas2021flexible}.

\begin{figure}[H]
	\centering
	\includegraphics[width=0.8\textwidth]{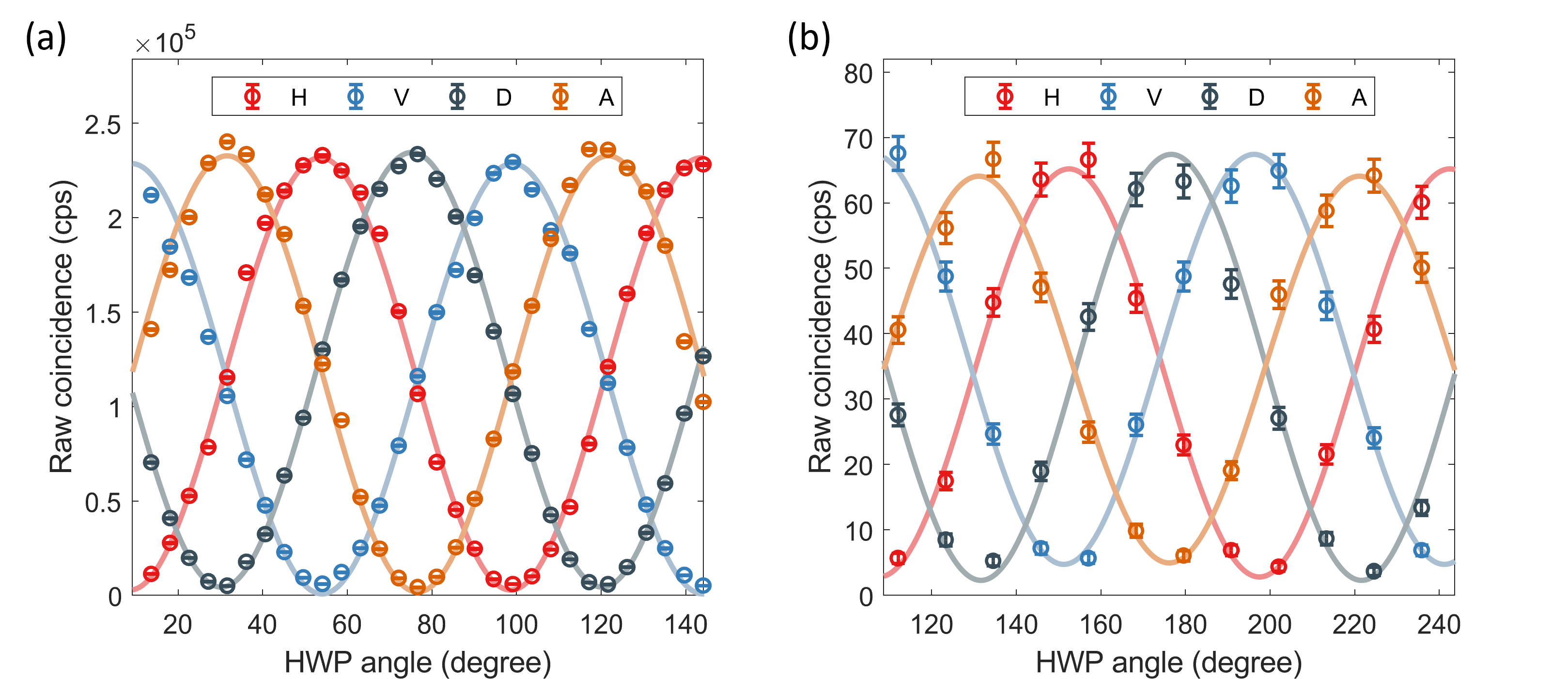}
		\caption{\textbf{Two fold coincidence curves.} (a) Coincidence measurement of entangled source. The signal photon's HWP angle $\theta_s$ is fixed at H, V, D, and A basis, while the idler photon's HWP angle $\theta_i$ rotates for \ang{180}. (b) Entanglement characterization with signal photons travel through 93~km of deployed fiber. Accidental counts 1.6~cps are not subtracted.}\label{93kmCurve}
\end{figure}

\section*{Entangled source characterization}\label{subsec1}
Once the signal and idler photons are filtered, they first pass through two quarter-wave plates (QWP) to compensate for polarization distortion caused by the connecting fibers. Each arm of the entangled photons are then projected into different polarization bases, using half-wave plates (HWP) and polarizing beam splitters. During source characterization, a short length of fiber replaces the deployed fibers, directly connecting the source to the measurement setup.

\begin{figure}[H]
	\centering
	\includegraphics[width=0.9\textwidth]{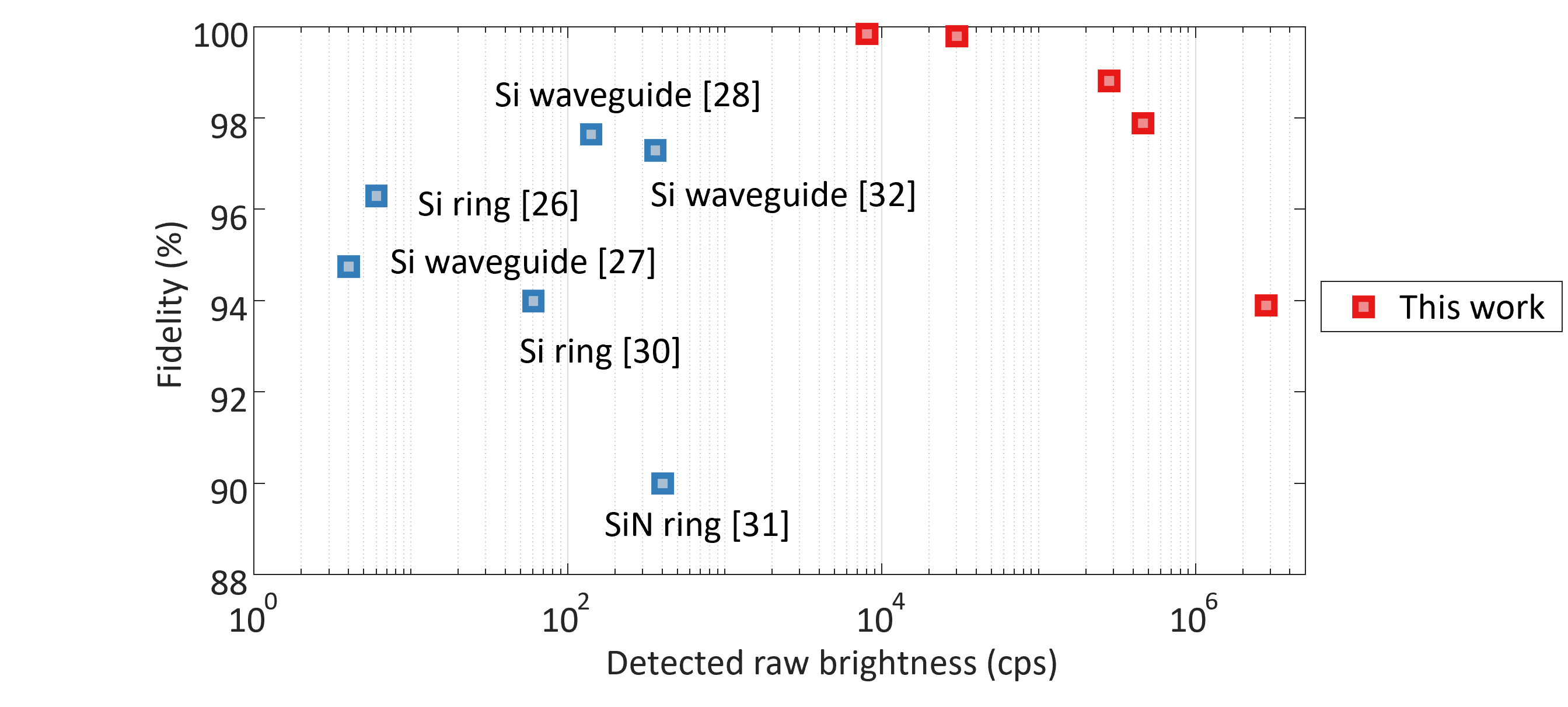}
	\caption{\textbf{Performance comparison of entangled sources.} The detected raw brightness and fidelity at different pump power are measured and compared the with reported data from other CMOS chip-based studies. Fidelity values are obtained from polarization visibility \cite{meyer2018high} with \mbox{\ensuremath{F=(1+V)/2}}. 
	The reduction in fidelity at higher brightness is primarily attributed to an increased probability of multi-photon emission at high pump power.}\label{literatureReview}
\end{figure}

The HWP in the signal arm is fixed at $\theta_s$, while the HWP in the idler arm rotates for \ang{180}. Photons are detected by SNSPDs, with the coincidence count recorded during the rotation. The single photons are detected with SNSPDs, with the efficiency for idler channel at 45\% while efficiency for signal channel is at 49\%.
Fig.~\ref{93kmCurve}.(a) displays the observed raw coincidence rate as a function of $\theta_i$, with the error bars estimated from Poisson statistics.
The maximum observed coincidence at the peak of the coincidence curves is 230 kcps, corresponding to an overall brightness of 460,000 entangled photon pairs a second. The fidelity to the bell state is 97.90(3)\%. This source is over 3 orders of magnitude brighter than sources in other literature reports. This comparison is provided in Fig.~\ref{literatureReview}.
When the source is operating at low brightness conditions, fidelity can be enhanced to 99.85(6)\%, and the experimental data detailed in Table~\ref{source comparison}.

To further verify the presence of entanglement, a Bell test was performed by measuring the violation of the Clauser-Horne-Shimony-Holt (CHSH) inequality. Polarization correlations in four measurement bases were recorded, and the CHSH $S$ parameter was calculated for the source at various brightness levels. The highest observed $S$ value is 2.82(2) at 8 kcps brightness, and $S$ remains well above the classical bound of 2 even at the highest measured pair rates.

\begin{table}[h]
	\centering
	\caption{Comparison with CMOS polarization entangled sources}
	\label{source comparison}
	\begin{tabular}{@{}cccccc@{}}
		\toprule
		Source type & Detector (eff.)  & Brightness (cps) & \quad Fidelity   & \quad  CHSH $S$  & Ref\\
		\midrule
		Si waveguide & APD (10\%) & 4 &\quad 94.8(41)\%&\quad - & \cite{takesue2008generation} \\
		Si waveguide & APD (20\%) & 140 &\quad 97.7(8)\% &\quad  2.7(1) & \cite{li2017chip} \\
		Si waveguide & SNSPD (85\%) & 360 &\quad 97.3(16)\% &\quad -  & \cite{zhang2019generation} \\
		Si ring & SPD (10\%) & 6 &\quad 96.3(15)\% &\quad -  & \cite{suo2015generation} \\
		Si ring & SNSPD (81\%) & 60 &\quad 94(2)\% &\quad -  & \cite{miloshevsky2024cmos} \\
		SiN ring & SNSPD (60\%) & 400 &\quad 90.0(-)\% &\quad -  & \cite{wen2023polarization} \\
		\hdashline
		\multirow{5}{*}{Si waveguide} & SNSPD (45\%, 49\%)& 8,000 &\quad 99.85(6)\%&\quad  2.82(2) & \multirow{5}{1cm}{\centering This \\work} \\
		& SNSPD (45\%, 49\%)& 30,000 &\quad 99.80(4)\% &\quad 2.815(8)  &\\
		& SNSPD (45\%, 49\%)& 280,000 &\quad 98.82(3)\% &\quad  2.777(3) & \\
		& SNSPD (45\%, 49\%)& 460,000 &\quad 97.90(3)\% &\quad 2.729(2) & \\
		& SNSPD (56\%)& 2,800,000* &\quad 93.9(7)\% &\quad 2.44(3) &\\
		\bottomrule
	\end{tabular}
	\noindent\begin{minipage}{\linewidth}
	\footnotesize{*This is measured with a 15 dB attenuator inserted in both the signal and idler arms before the SNSPDs to avoid detector saturation. The decreasing fidelity with increasing brightness is due to the number of multiple photon pairs present within the coincidence time window of this system. This effect can be further minimized if the system timing jitter and coincidence window is reduced.}
	\end{minipage}
\end{table}

\section*{Asymmetric entanglement distribution over 93 km distance}\label{subsec1}
The signal photons are first connected to the NUS-SUTD-NUS deployed fiber which is 93 km, exhibiting a 35.4 dB loss as measured by an optical time domain reflectometer (OTDR). The idler photons are detected at the source. 
The propagation loss is calculated to be 0.38 dB/km, which is higher than the typical 0.16 dB/km observed in fiber spools or 0.25 dB/km in submarine fibers \cite{wengerowsky2019entanglement,wengerowsky2020passively}. This increased loss is primarily due to the connection junctions in practical metropolitan fibers. When the dispersion compensation module (DCM) is added, 4.3 dB extra loss is also included to the fiber system.

To evaluate the entanglement quality after transmission through the NUS-SUTD-NUS fiber loop, we perform two-fold coincidence measurements equivalent to local testing. The idler arm's polarization analysis setup is set to H, V, D, and A bases. The signal arm's HWP is rotated through \ang{180} to measure the visibility. As shown in Fig.~\ref{93kmCurve}.(b), observed raw count rate of entangled photons is 132 cps, while the fidelity is 93.3(3)\% in the 10 second measurement duration. The high fidelity attributed to effective background count filtering and nonlocal dispersion compensation. 

\section*{Near-symmetric entanglement distribution over 155 km deployed fiber}\label{subsec1}
When only the signal photons are deployed, the timing delay between the signal and idler arms are observed to experience a cyclic drift. This cyclic drift is attributed to diurnal temperature fluctuations that cause the length of the fibers to expand and contract. This effect is reduced when both signal and idler photons are deployed over similar lengths of fiber.
This ``near-symmetric" experiment was performed by connecting the idler photons to a fiber loop that connected NUS and NTU. In this setup, the overall length traversed by the photon pair exceeds 155 km, and the overall system loss reached 66 dB (56 dB from fiber, 10 dB from 160~km version DCM). Although the final coincidence detection is performed locally at NUS, both the signal and idler photons are independently transmitted over two separate long-distance fiber link before measurement. This effectively simulates a distributed entanglement scenario, as the entangled photons travel through metropolitan deployed fibers and experience actual losses, dispersion, and changes from the environment. By increasing the pump power, the entangled photon pair rate after transmission was 0.7 cps with a fidelity of 87.6(5)\%. Accounting for fiber loss, the off-chip photon pair rate was 2.8 Mcps and the estimated fidelity was 93.9(7)\% as shown in Fig.~\ref{literatureReview}. The degradation is primarily due to the effects of polarization mode dispersion (PMD) \cite{neumann2022continuous,gisin2002quantum,shi2020stable,zhang2024polarizationencodedquantumkeydistribution,antonelli2011sudden}.
A comparison of the results from this work with similar work in the literature regarding fiber-based entanglement distribution is provided in Table~\ref{long distance comparison}.

\begin{table} [h]
	\centering
	\caption{Comparison with other long fiber works}\label{long distance comparison}
	\begin{tabular}{@{}cccccccc@{}}
		\toprule
		Source type & Fiber type & Fiber loss & DOF & Rate (cps) & Fidelity&  Ref\\
		\midrule
		\makecell{Sagnac\\PPLN} & \makecell{Submarine \\ (96 km)} & \makecell{22 dB} & pol. & 366 & 95.3(3)\% & \cite{wengerowsky2019entanglement}\\
		\makecell{Sagnac\\PPLN} & \makecell{Submarine \\ (192 km)} & \makecell{48 dB} & pol. & 8.6 & 95(1)\%   &\cite{wengerowsky2020passively}\\
		\makecell{Sagnac\\PPLN} & \makecell{Metropolitan \\ (248 km)} & \makecell{71.5 dB} & pol. & 9 & 93(-)\%&\cite{neumann2022continuous}\\
		\makecell{Sagnac\\PPLN} & \makecell{Ultra low loss spool \\ (404 km)} & 87 dB & pol. & 0.03 &91.3(-)\%  &\cite{zhuang2024ultrabright}\\
		\hdashline
		\multirow{3}{*}{\makecell{Sagnac\\Si chip}}& \makecell{Metropolitan \\ (93 km)} & \makecell{39.7 dB} & \multirow{3}{*}{pol.} & 132 & 93.3(3)\%&  \multirow{3}{1cm}{\centering This \\ work}\\
		&\makecell{Metropolitan \\ (155 km)}&\makecell{66 dB}& &0.7&87.6(5)\%  &\\
		
		\bottomrule
		
	\end{tabular}
\end{table}
\section*{Discussion}\label{sec3}

In this report we have demonstrated a bright entangled photon pair source based on a silicon nanophotonic chip. 
Through system-level optimization, the source enabled polarization entanglement distribution over 155~km of deployed fiber (66~dB loss).
The observed brightness and fidelity of the nanophotonic chip source is competitive when compared against work based on bulk crystal sources \cite{wengerowsky2019entanglement,wengerowsky2020passively,neumann2022continuous}.
This work marks a milestone in the capability of silicon nanophotonic chips for enabling long range quantum networks. 

One important challenge in distributing polarization-entangled photons over long fibers is PMD, which can degrade entanglement quality. 
In addition to recently proposed compensation techniques \cite{xingjian2025polarization,rodimin2024impact}, microring resonators (MRRs) offer an alternative approach by generating narrowband photon pairs that are inherently less susceptible to PMD.

The system level optimization concepts proposed by this paper can be applied to other integrated photonics platforms like AlGaAs \cite{steiner2023continuous,autebert2016integrated,appas2021flexible,steiner2021ultrabright}, TFLN \cite{shi2024efficient,zhao2020high}, SiC \cite{lohrmann2017review,rahmouni2024entangled}, USRN \cite{choi2020correlated}, AlN \cite{guo2017parametric}.  
This source is also useful for satellite to ground experiments because it exceeds the loss of the Micius double-downlink where the link loss was 62 dB \cite{yin2017satellite,lu2022micius}.

There are a number of avenues for further integration and improvement. For example, many of the filters are based on fiber components and further integration \cite{ong2013ultra,luo2014wdm,kumar2020compact} is possible in this aspect enhancing the stability and scalability of the setup. The nonlocal dispersion compensation is performed using a dispersion compensated module that has a significant transmission loss, which should be reduced with further engineering \cite{cui2022programmable,ong2022dispersion}. Additionally, there is a huge potential for utilising dense-wavelength division multiplexing with the silicon nanophotonic chip. This will expand the number of channels accessible \cite{zhuang2024ultrabright,joshi2018frequency,miloshevsky2024cmos}. Finally, other degrees of freedom, such as the time-bins \cite{xiong2015compact,liu2019energy,liu2023high,wen2022realizing,lu2020three} could be added to the setup to enable high-dimensional entanglement distribution \cite{zheng2023multichip,xie2015harnessing,chapman2020time}.
Taken together, these strategies can significantly enhance the quantum information capacity of integrated photonic platforms and accelerate the development of scalable, long-range quantum networks.

We note that a preprint of a related study \cite{jiang2025entanglement} was posted in March 2025, six months after posting the current work on arXiv.

\section*{METHODS}


\subsection*{Photon pair wavelength selection}\label{subsec3}
In the SFWM process, two pump photons are absorbed simultaneously, elevating the system to a virtual energy level, followed by the emission of a signal and idler photon to conserve both energy and momentum. The efficiency of this nonlinear interaction is highly dependent on the phase matching conditions. 
The coincidence-to-accidentals ratio (CAR) quantifies the ratio of true photon coincidences (from genuine photon pairs) to accidental coincidences, such as those arising from multi-photon emissions or system noise. A high CAR is essential for ensuring high entanglement fidelity, as it directly correlates with the purity of the detected entangled state by reducing the likelihood of accidental events degrading the measurement. Therefore, maximizing CAR requires lower multi-photon events and system noise, such as residual pump photons and Raman noise.  

To select the optimal wavelengths for the highest CAR and entangled photon quality, a 1550.12 nm CW laser is employed to unidirectional pumping for the same chip used in the entanglement generation experiment.
The generated photon pairs, signal and idler, were spectrally broad due to the nature of phase matching in silicon. Two 0.8 nm bandwidth filters with central wavelength symmetrical about the pump are installed to scan the pair generation rate across different wavelength channels.

\begin{figure}[H]
	\centering
	\includegraphics[width=0.8\textwidth]{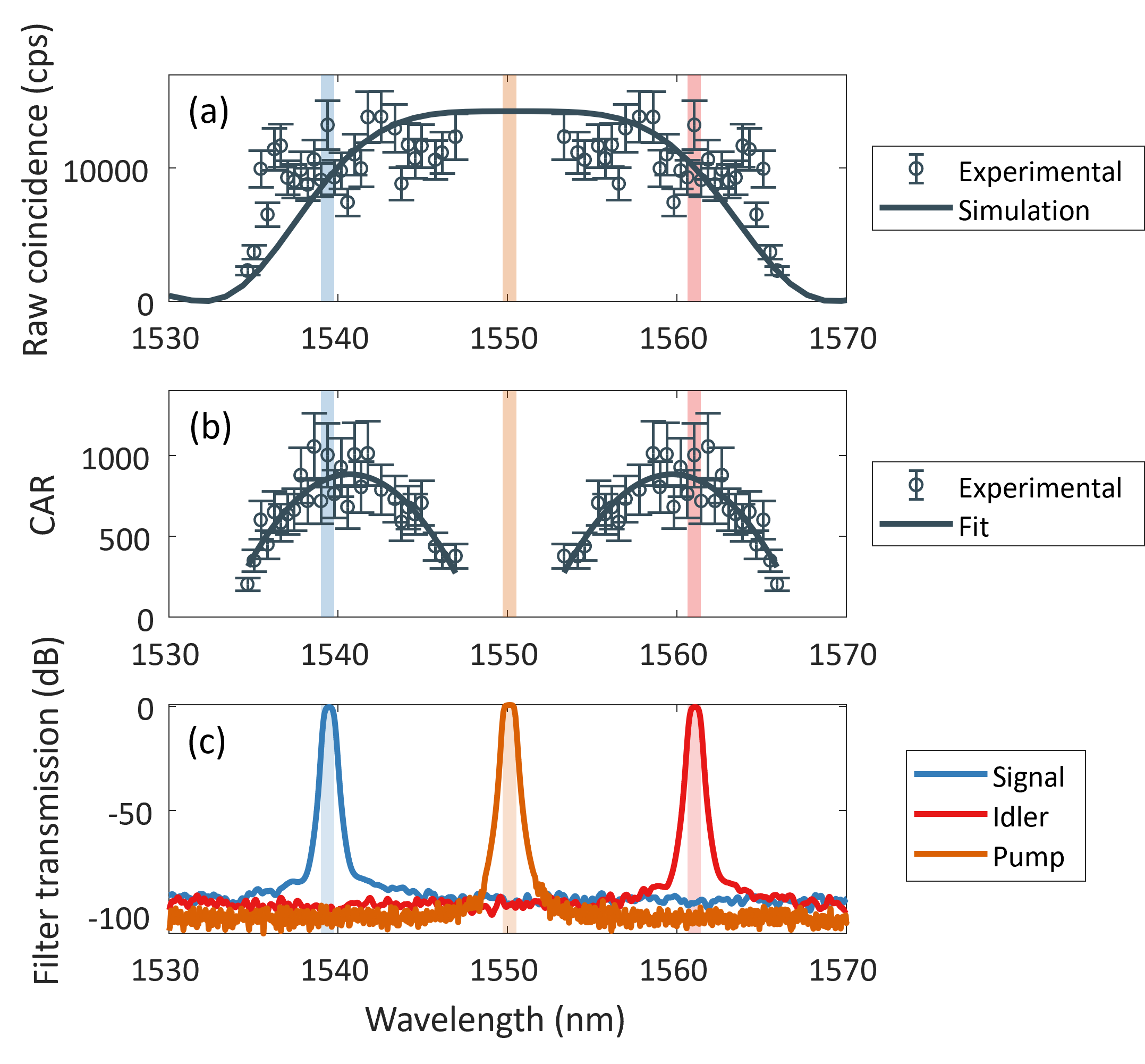}
	\caption{\textbf{Observed photon pair rate, CAR, and filter transmission as functions of wavelength.} (a) Raw coincidence rate as a function of signal and idler wavelengths relative to the pump, showing that photon pair generation is most efficient near the pump wavelength due to optimal phase matching. (b) Coincidence-to-accidental ratio (CAR) as a function of wavelength. The CAR peaks approximately 10 nm away from the pump, where pump photon rejection is optimal. (c) Transmission profiles of the filters used for signal, idler, and pump photons, with an extinction ratio of approximately 80-100 dB, indicating the optimal filter performance at a 5-10 nm separation from the center wavelength.
	The shaded regions in panels (a-c) correspond to the selected passbands of the three filters employed in the final entanglement distribution experiment.}
	\label{selectwl}
\end{figure}
As illustrated in Fig.~\ref{selectwl}.(a), the raw photon coincidence rate was highest when the signal and idler wavelengths were closest to the pump wavelength, indicating that phase matching was most efficient at shorter detunings from the pump. However, this increased photon pair generation does not directly translate into a higher CAR, as shown in  Fig.~\ref{selectwl}.(b), where the CAR peaks approximately 10~nm away from the pump wavelength. 
This discrepancy can be explained by the balance between photon pair generation and system noise. The error bars in Fig.~\ref{selectwl}(a,b) represent uncertainties estimated from Poisson statistics and DWDM connector insertion loss variations. While photon pairs are generated more efficiently near the pump wavelength, the proximity of the signal and idler to the pump results in less effective rejection of residual pump photons as shown in Fig.~\ref{selectwl}.(c). These residual pump photons contribute to accidental coincidences, lowering the CAR. On the other hand, when the signal and idler photons are too far from the pump, the SFWM process itself becomes less efficient, reducing the photon pair generation rate and thereby lowering the CAR due to the lower coincidence rate. The observed peak in CAR at a 10 nm separation from the pump wavelength results from the optimal trade-off between reducing system noise and maintaining sufficient photon pair generation.

\subsection*{Chip design and coupling strategy}\label{subsec3}

\begin{figure}[h]
	\centering
	\includegraphics[width=0.8\textwidth]{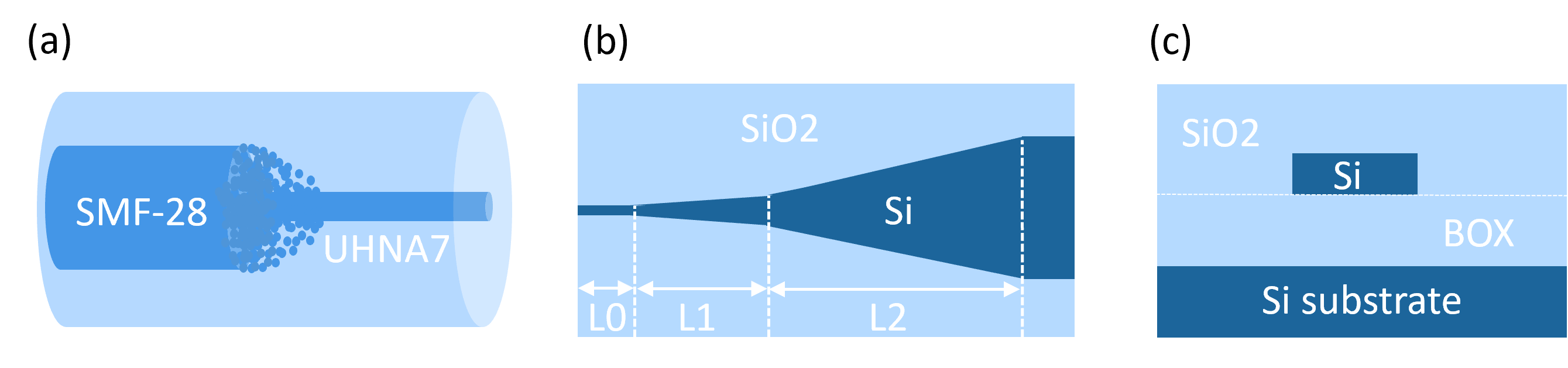}
	\caption{\textbf{Chip coupling strategy.} (a) Fiber splicing junction showing the tapered region that adiabatically converts the UHNA7 fiber mode to SMF-28. (b) Top view of the two-stage inverse taper edge coupler with a BOX layer; upper cladding not shown. (c) Cross-sectional view of a silicon chip with a silicon waveguide above a \SI{2}{\micro\meter} BOX layer, covered by \SI{2}{\micro\meter} SiO2 cladding.}
	\label{chip}
\end{figure}
The chip involved in the experiment is a 8 mm long silicon waveguide with cross sectional dimension 650 nm in width and 250 nm in height. The whole chip is covered by \SI{2}{\micro\meter} thick SiO2 cladding layer. 
On the chip edges, two-stage inverse tapers are fabricated to expand the chip mode to reduce the mode mismatch with that of the UHNA7 fiber as shown in Fig.~\ref{chip}. The L0 section acts as buffer region to protect the edge coupler during dicing process, while the length of L1 and L2 tapers are \SI{50}{\micro\meter} and \SI{100}{\micro\meter} respectively. Short section of UHNA7 fiber (less than 1 cm) is spliced to SMF-28 fiber to assist light collection. The loss is optimized to 0.03 dB per splicing junction by performing an in-situ monitoring of the loss during splicing process \cite{du2024demonstration,yin2019low}. The measured coupling loss is 0.64 dB per facet and has nice long term stability of \(\pm 0.02\) dB for more than ten days without using any adhesive. 

\begin{figure}[H]
	\centering
	\includegraphics[width=0.8\textwidth]{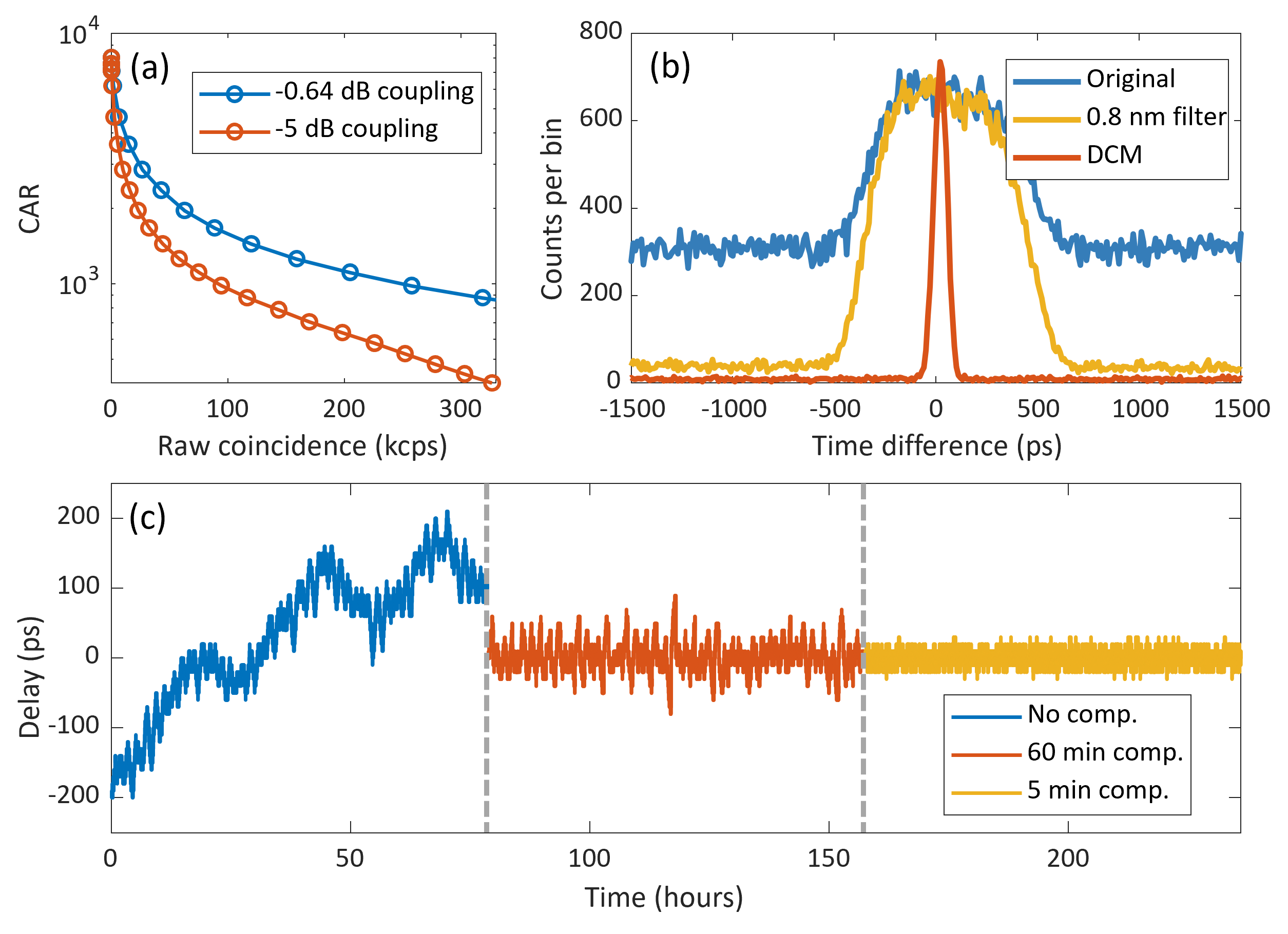}
	\caption{\textbf{Methods for improving entanglement fidelity in deployed fibers.} (a) The degradation of CAR due to coupling loss is more pronounced at higher observed brightness. The 0.64~dB loss case achieves a CAR more than two times higher than that of the 5~dB coupling loss case. (b) The peaks are in the center of the cross-correlation function between the time tag files obtained from the two SNSPD channels. The time delay \SI{457}{\micro\second} between the two channels is already compensated digitally. The blue line is the original histogram shape after the long fiber transmission. The green line is the histogram with dark count noise filtering. After nonlocal dispersion compensation, the FWHM of the histogram peak is greatly reduced and plotted in red line. (c) Measured delay drift without compensation (blue), with 60-minute feedback (orange), and with 5-minute feedback (yellow). The 5-minute compensation confines the delay within \(\pm 20~\text{ps}\) for 99.6\% of the time.}
	\label{CARVsLossHistDridt}
\end{figure}

Low coupling loss is critical for practical entangled photon-pair sources, not only because it enhances the overall brightness, but also because it significantly impacts the CAR \cite{harada2009frequency}, as shown in Fig.~\ref{CARVsLossHistDridt}.(a). 
The relationship between the CAR, the on-chip raw coincidence \(N_c\), and the system efficiencies \(\alpha\) can be expressed as:
\begin{align}
	\mathrm{CAR} = \frac{N_c \cdot \alpha^2}{\bigl[(N_c + N_{\mathrm{noise}})\alpha + N_{\mathrm{dark}}\bigr]^2} + 1,
	\label{eq:CAR_formula}
\end{align}

In SFWM, the pair generation rate scales quadratically with pump power, \(N_c \propto P^2\) while the background noise (Raman scattering and unfiltered pump) scales linearly \(N_{\mathrm{noise}} \propto P\). For a fixed detected coincidence rate \ensuremath{N_c \alpha^2 = \mathrm{const}}, it follows that \(N_c \propto 1/\alpha^2\), \(P \propto 1/\alpha\), and hence \(N_{\mathrm{noise}} \propto 1/\alpha\). The denominator of Eq.~\ref{eq:CAR_formula} thus scales as
\begin{align}
	\bigl[(N_c + N_{\mathrm{noise}})\alpha + N_{\mathrm{dark}}\bigr]^2 \sim \bigl(1/\alpha + \mathrm{const}\bigr)^2,
\end{align}
indicating that CAR increases monotonically with coupling efficiency \(\alpha\).

\subsection*{Nonlocal dispersion compensation}\label{nonlocal dispersion compensation}
In this experiment the deployed fiber is not dark and contains broadband noise within the optical channel of interest. This noise lowers the CAR value affecting the fidelity of the observed quantum state. The maximum achievable fidelity is limited by the CAR and is given by $F = \frac{CAR}{CAR+1}$.
A 0.8 nm filter is installed before the SNSPDs, and the dark count noises are effectively removed from around 4 Mcps to less than 200 cps. As shown in Fig.~\ref{CARVsLossHistDridt}.(b), the CAR is increased from 2 to 15, which corresponds to a fidelity upper bound improvement from 67\% to 94\%.

Additionally, chromatic dispersion presents a significant challenge to achieve high entanglement fidelity. 
The dispersion value at 1539 nm is around $D_{1539} \leq 18.0$ ps/(nm\(\cdot\)km) in the fiber. Since the bandwidth of signal photons is 0.8 nm while the fiber distance is 93 km, the dispersion caused by the deployed fiber is 1339 ps. 
In comparison, the overall timing jitter of the SNSPDs including the time tagger, is approximately 80 ps.
As a result, measuring the photon pair rate after signal photons transmitted through the long fiber requires a larger coincidence window around 1 ns. This larger coincidence window increases accidental counts, greatly reducing the CAR value. 
To solve this, a dispersion compensation module (DCM) is inserted in the idler arm. The total dispersion compensation provided by the DCM at 1561 nm is around -1360 ps/nm, as data provided by the manufacturer. 
As shown in Fig.~\ref{CARVsLossHistDridt}.(b), effective dispersion compensation raises the CAR from 15 to 91, corresponding to an increase in theoretical fidelity from 94\% to 99\%. However, the measured fidelity in practice is lower than this theoretical value, primarily due to the effects of PMD. It is important to note that the DCM introduces 4.3 dB insertion loss, which does impact the coincidence rate. This DCM-related loss may be mitigated in future work where the waveguide platform could be designed to generate photons with narrower wavelength bandwidth \cite{ma2016silicon,suo2015generation,miloshevsky2024cmos,wen2023polarization,zhang2024polarization}.

\subsection*{Phase stabilization in deployed fibers}\label{phaseStab}
 In the 93 km asymmetric configuration, phase fluctuations are more pronounced because only one photon passes through the long fiber while the other is detected locally. In this case, environmental changes such as temperature-induced fiber expansion or contraction directly translate into phase drift between the two optical paths.
In contrast, the near-symmetric entanglement distribution experiment utilizes both the 93 km and 62 km deployed fiber segments, with each photon of the entangled pair traversing one segment. This configuration benefits from partial passive phase noise cancellation: since both fibers are deployed underground in similar metropolitan conditions, environmental fluctuations tend to affect both arms similarly, leading to correlated phase changes and reduced differential phase drift. 
The phase stability experiments were performed on the 62 km fiber segment, and the results are shown in Fig.~\ref{CARVsLossHistDridt}.(c). The observed delay drift in this configuration is representative of the deployed metropolitan fiber environment.
Every 5 minutes, the Sagnac pump splitting ratio is switched from 1:1 to 1:0 by rotating the HWP in the pump polarization control unit in Fig.~\ref{setup}.(a), and the waveplates in the state analyzer  in Fig.~\ref{setup}.(e) are adjusted to measure the HH coincidence histogram. This temporary configuration doubles the pump power and hereby increases the coincidence count by a factor of four, enhancing the precision of delay estimation. The delay is extracted by fitting the coincidence histogram with a Gaussian envelope and summing five bins around the peak (when using 50~ps coincidence window), with the central bin identified as the reference delay. During the subsequent 5-minute measurement period, the pump splitting ratio is reverted to 1:1, and the previously extracted delay value is applied. This calibration cycle is repeated at 5-minute intervals.

As shown in Fig.~\ref{CARVsLossHistDridt}.(c), the uncompensated phase drift can exceed \(\pm 200~\text{ps}\) over time. A 60-minute feedback interval provides moderate suppression, whereas a 5-minute compensation cycle confines the delay variation within \(\pm 20~\text{ps}\) for more than 70 hours, achieving a stabilization accuracy of 99.6\%.
These results highlight the importance of employing high-brightness entangled photon-pair sources. Without sufficient brightness, longer integration times are required to obtain statistically significant data, during which uncorrected phase drift can significantly degrade the CAR and reduce the fidelity of the reconstructed quantum state.

\subsection*{System calibration}\label{subsec3}
The polarization shift caused by birefringence is compensated by the QWP and HWP in the measurement setup. To calibrate the system once the NUS-NTU-NUS quantum channel is established, we disconnect the H polarization arm of Sagnac loop and inject a 13 dBm laser beam at 1561.01 nm into the setup. This wavelength matches that of the idler photons, allowing the laser to go through the same route as idler photons do. After transmission through the PBS in the measurement setup, the laser is detected by a free space powermeter. The QWP and HWP are then adjusted to minimize the detected power, with the final HWP angle designated as the V state for the idler arm. The output power difference between HWP's H and V position is 37 dB which is limited by the extinction ratio of the polarization components. A similar procedure is employed for the signal arm, using a laser at 1539.37 nm. Once the waveplates' angles are calibrated, it is crucial that the QWPs remain stationary during the two-photon interference measurement to ensure accuracy.
In the local measurement case, once the photon pairs are collected with the UHNA7 fiber, the insertion loss from Sagnac loop to the fiber connector before SNSPD is 2.74 dB for signal arm and 2.72 dB for idler arm. This includes all the fiber based and free space optical components in the experiment.

\subsection*{Single photon detection and coincidence counting}\label{subsec3}
The entangled photons are detected by SNSPDs. The detection efficiency of the SNSPDs is tunable via the device bias current: higher bias increases both detection efficiency and dark count rate, but can reduce the maximum detectable count rate by making the detector more susceptible to latching.

For source characterization measurements, the SNSPD efficiency was intentionally set to a low value (45\% and 49\%) to increase the pump power substantially without saturating the detectors.

In the 93 km entanglement distribution experiment, the single photons were detected with SNSPDs set to higher efficiency: 78\% for the signal channel and 45\% for the idler channel.  The single-photon count rates were 4.8 kcps for the signal arm and 5.6 Mcps for the idler arm. The delay caused by the deployed fiber is 457,369,970 ps, matching the distance measured by OTDR. The coincidence window $\tau_{cw}$ used in local measurements is 200 ps, while $\tau_{cw}$ with deployed fiber is 60 ps for accidental suppression and higher CAR. The narrow $\tau_{cw}$ results in a lower photon-pair rate. Reducing the background noise (detector dark counts, Raman noise in the source, and stray light coupling into the fiber network) should further improve the CAR and fidelity.

When the deployed fiber distance was extended to 155 km, the singles count rates on the signal and idler arms were around 15 kcps and 40 kcps, respectively. Although this low single-photon count rate does not saturate the detector, the SNSPDs were set to 56\% efficiency by lowering the bias current to reduce the dark count rate to 10 cps. The optical delay caused by the 62 km fiber is 302,113,173 ps, consistent with the length characterization by OTDR.

\begin{figure}[H]
	\centering
	\includegraphics[width=0.8\textwidth]{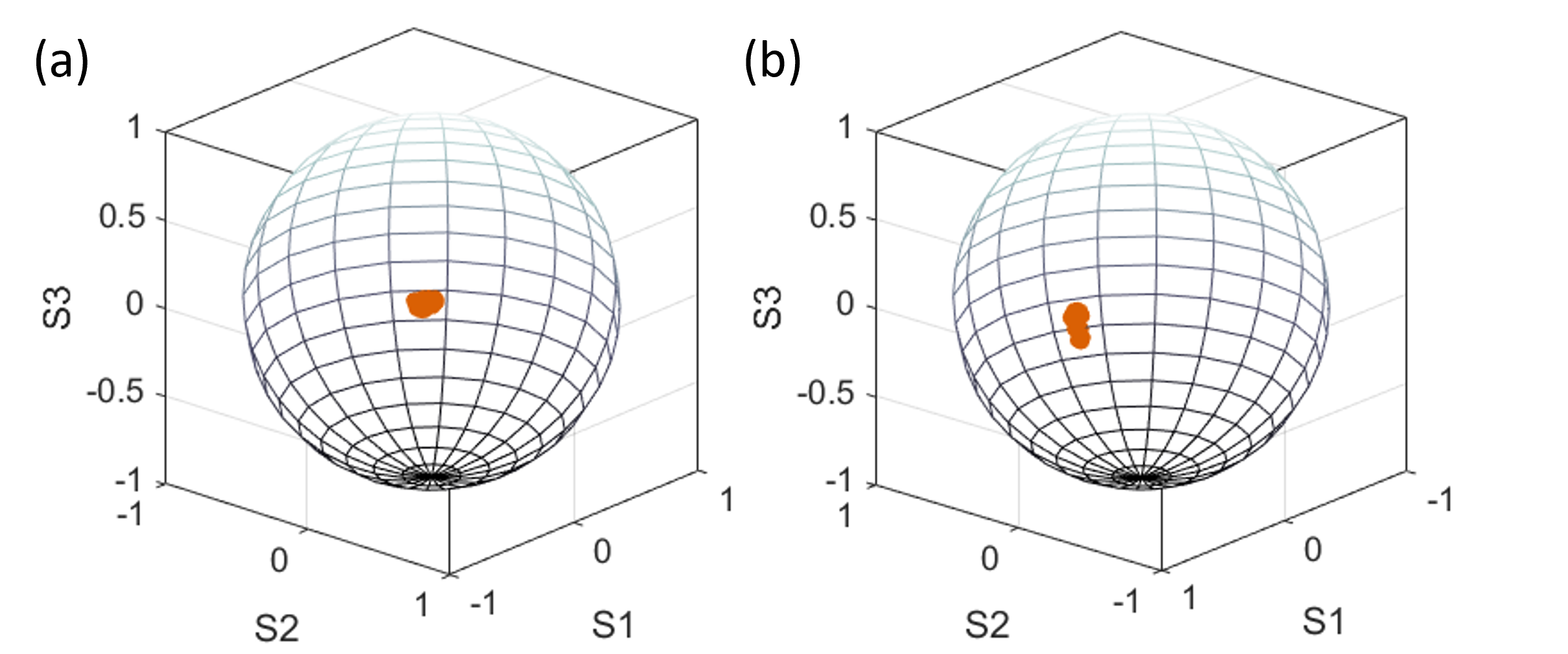}
	\caption{\textbf{Long-term polarization stability of deployed fiber links.}
		Polarized light at 1539.37~nm and 1561.01~nm is transmitted through (a) the 93~km NUS-SUTD-NUS link and (b) the 62~km NUS-NTU-NUS link.
		The output polarization states are continuously monitored over 8~hours and plotted on the Poincaré sphere, where the three axes $S_1$, $S_2$, and $S_3$ correspond to the normalized Stokes parameters. Each orange cluster represents the temporal evolution of the output polarization state, indicating minor polarization drift observed over time in both deployed fibers.}
	\label{PMD}
\end{figure}

\subsection*{Fiber link information}\label{subsec3}

A measurement of polarization stability of deployed fibers over a continuous period of 8 hours is conducted and the polarization state rotates \ang{10.75} (93 km link) and \ang{12.63} (62 km link) on the Poincaré sphere. The results in Fig.~\ref{PMD}. (a, b) indicate a relatively stable polarization behavior. 

Additionally, signal crosstalk from other channels in metropolitan fiber networks can introduce background noise, which also affects the quantum state fidelity. The NUS-SUTD-NUS (or NUS-NTU-NUS) fiber loop experiences approximately 4 Mcps (2 kcps) of background noise without spectral filtering. This noise is broadband and can be effectively mitigated using a 0.8 nm bandwidth DWDM filter. After inserting the filter, the background noise in both channels is reduced to less than 200 cps.

\newpage


\section*{RESOURCE AVAILABILITY}


\subsection*{Lead contact}


Requests for further information and resources should be directed to and will be fulfilled by the lead contact Alexander Ling (alexander.ling@nus.edu.sg).

\subsection*{Materials Availability}
This study did not generate new unique materials.
\subsection*{Data and code availability}


\begin{itemize}
	\item The data reported in this study are available from the lead author upon reasonable request.
	\item This paper does not report any original code.
	\item Any additional information required to reanalyze the data reported in this paper is available from the lead contact upon request.
\end{itemize}

\section*{ACKNOWLEDGMENTS}

This research is supported by the National Research Foundation, Singapore, and A*STAR under its CQT Bridging Grant and Quantum Engineering Programme (Award No. NRF2021-QEP2-01-P02), QEP2 grant: NRF2021-QEP2-01-P01, and US Air Force grant (FA2386-21-1-0076). 
D. Tan acknowledges funding from the National Research Foundation,
Singapore, and A*STAR under its QEP 2.0 Programme (Award No. NRF2022-QEP2-01-P08), the National Research Foundation Investigatorship (NRF-NRFI08-2022-0003) and the Ministry of Education Tier 2 Grant (Award No. MOE2019-T2-2-178).
We acknowledge funding support from the National Research Foundation, Singapore and A*STAR under its Quantum Engineering Programme (National Quantum-Safe Network, NRF2021-QEP2-04-P01) and Netlink Trust for the provisioning of the fiber network. The authors acknowledge the assistance of setup alignment and literature review from Arya Chowdhury, and En Teng Lim, and consultation from Dr.~Tanvirul Islam, Dr.~Haoran Zhang, and Dr.~Shihao Ru. 

\section*{AUTHOR CONTRIBUTIONS}
J.D., X.Z., D.T.H.T., and A.L. conceived the experiment;
G.F.R.C. and H.G. contributed to materials processing and fabrication;
J.D. performed chip characterization; 
J.D. and X.Z. conducted entanglement measurements and analyzed the data;
J.D., D.T.H.T., and A.L. wrote the manuscript;
A.L. and D.T.H.T. supervised the project and acquired funding.

\section*{DECLARATION OF INTERESTS}

Information related to the edge coupler design used in this work is covered by a pending international patent application (Application No. PCT/SG2024/050518, titled “Coupling Platform for Photonic Chips”) filed by Jinyi Du, Dawn T. H. Tan, and Alexander Ling.

\bibliography{references}

\begin{thebibliography}{69}
\providecommand{\natexlab}[1]{#1}
\providecommand{\url}[1]{\texttt{#1}}
\providecommand{\href}[2]{#2}
\providecommand{\path}[1]{#1}
\providecommand{\DOIprefix}{doi: }
\providecommand{\ArXivprefix}{arXiv: }
\providecommand{\URLprefix}{URL: }
\providecommand{\Pubmedprefix}{pmid: }
\providecommand{\doi}[1]{\href{http://dx.doi.org/#1}{\path{#1}}}
\providecommand{\Pubmed}[1]{\href{pmid:#1}{\path{#1}}}
\providecommand{\BIBand}{and}
\providecommand{\bibinfo}[2]{#2}
\ifx\xfnm\undefined \def\xfnm[#1]{\unskip,\space#1}\fi
\makeatletter\def\@biblabel#1{#1.}\makeatother
\bibitem[{Liao et~al.(2017)Liao, Cai, Liu, Zhang, Li, Ren, Yin, Shen, Cao, Li
  et~al.}]{liao2017satellite}
\bibinfo{author}{Liao, S.K.}, \bibinfo{author}{Cai, W.Q.},
  \bibinfo{author}{Liu, W.Y.}, \bibinfo{author}{Zhang, L.},
  \bibinfo{author}{Li, Y.}, \bibinfo{author}{Ren, J.G.}, \bibinfo{author}{Yin,
  J.}, \bibinfo{author}{Shen, Q.}, \bibinfo{author}{Cao, Y.},
  \bibinfo{author}{Li, Z.P.} et~al. (\bibinfo{year}{2017}).
  \bibinfo{title}{Satellite-to-ground quantum key distribution}.
\newblock \bibinfo{journal}{Nature} \emph{\bibinfo{volume}{549}},
  \bibinfo{pages}{43--47}.
\bibitem[{Yin et~al.(2017)Yin, Cao, Li, Ren, Liao, Zhang, Cai, Liu, Li, Dai
  et~al.}]{yin2017satellite}
\bibinfo{author}{Yin, J.}, \bibinfo{author}{Cao, Y.}, \bibinfo{author}{Li,
  Y.H.}, \bibinfo{author}{Ren, J.G.}, \bibinfo{author}{Liao, S.K.},
  \bibinfo{author}{Zhang, L.}, \bibinfo{author}{Cai, W.Q.},
  \bibinfo{author}{Liu, W.Y.}, \bibinfo{author}{Li, B.}, \bibinfo{author}{Dai,
  H.} et~al. (\bibinfo{year}{2017}). \bibinfo{title}{Satellite-to-ground
  entanglement-based quantum key distribution}.
\newblock \bibinfo{journal}{Physical review letters}
  \emph{\bibinfo{volume}{119}}, \bibinfo{pages}{200501}.
\bibitem[{Villar et~al.(2020)Villar, Lohrmann, Bai, Vergoossen, Bedington,
  Perumangatt, Lim, Islam, Reezwana, Tang et~al.}]{villar2020entanglement}
\bibinfo{author}{Villar, A.}, \bibinfo{author}{Lohrmann, A.},
  \bibinfo{author}{Bai, X.}, \bibinfo{author}{Vergoossen, T.},
  \bibinfo{author}{Bedington, R.}, \bibinfo{author}{Perumangatt, C.},
  \bibinfo{author}{Lim, H.Y.}, \bibinfo{author}{Islam, T.},
  \bibinfo{author}{Reezwana, A.}, \bibinfo{author}{Tang, Z.} et~al.
  (\bibinfo{year}{2020}). \bibinfo{title}{Entanglement demonstration on board a
  nano-satellite}.
\newblock \bibinfo{journal}{Optica} \emph{\bibinfo{volume}{7}},
  \bibinfo{pages}{734--737}.
\bibitem[{Yin et~al.(2020)Yin, Li, Liao, Yang, Cao, Zhang, Ren, Cai, Liu, Li
  et~al.}]{yin2020entanglement}
\bibinfo{author}{Yin, J.}, \bibinfo{author}{Li, Y.H.}, \bibinfo{author}{Liao,
  S.K.}, \bibinfo{author}{Yang, M.}, \bibinfo{author}{Cao, Y.},
  \bibinfo{author}{Zhang, L.}, \bibinfo{author}{Ren, J.G.},
  \bibinfo{author}{Cai, W.Q.}, \bibinfo{author}{Liu, W.Y.},
  \bibinfo{author}{Li, S.L.} et~al. (\bibinfo{year}{2020}).
  \bibinfo{title}{Entanglement-based secure quantum cryptography over 1,120
  kilometres}.
\newblock \bibinfo{journal}{Nature} \emph{\bibinfo{volume}{582}},
  \bibinfo{pages}{501--505}.
\bibitem[{Ursin et~al.(2007)Ursin, Tiefenbacher, Schmitt-Manderbach, Weier,
  Scheidl, Lindenthal, Blauensteiner, Jennewein, Perdigues, Trojek
  et~al.}]{ursin2007entanglement}
\bibinfo{author}{Ursin, R.}, \bibinfo{author}{Tiefenbacher, F.},
  \bibinfo{author}{Schmitt-Manderbach, T.}, \bibinfo{author}{Weier, H.},
  \bibinfo{author}{Scheidl, T.}, \bibinfo{author}{Lindenthal, M.},
  \bibinfo{author}{Blauensteiner, B.}, \bibinfo{author}{Jennewein, T.},
  \bibinfo{author}{Perdigues, J.}, \bibinfo{author}{Trojek, P.} et~al.
  (\bibinfo{year}{2007}). \bibinfo{title}{Entanglement-based quantum
  communication over 144 km}.
\newblock \bibinfo{journal}{Nature physics} \emph{\bibinfo{volume}{3}},
  \bibinfo{pages}{481--486}.
\bibitem[{Yin et~al.(2012)Yin, Ren, Lu, Cao, Yong, Wu, Liu, Liao, Zhou, Jiang
  et~al.}]{yin2012quantum}
\bibinfo{author}{Yin, J.}, \bibinfo{author}{Ren, J.G.}, \bibinfo{author}{Lu,
  H.}, \bibinfo{author}{Cao, Y.}, \bibinfo{author}{Yong, H.L.},
  \bibinfo{author}{Wu, Y.P.}, \bibinfo{author}{Liu, C.}, \bibinfo{author}{Liao,
  S.K.}, \bibinfo{author}{Zhou, F.}, \bibinfo{author}{Jiang, Y.} et~al.
  (\bibinfo{year}{2012}). \bibinfo{title}{Quantum teleportation and
  entanglement distribution over 100-kilometre free-space channels}.
\newblock \bibinfo{journal}{Nature} \emph{\bibinfo{volume}{488}},
  \bibinfo{pages}{185--188}.
\bibitem[{Marcikic et~al.(2004)Marcikic, De~Riedmatten, Tittel, Zbinden,
  Legr{\'e} and Gisin}]{marcikic2004distribution}
\bibinfo{author}{Marcikic, I.}, \bibinfo{author}{De~Riedmatten, H.},
  \bibinfo{author}{Tittel, W.}, \bibinfo{author}{Zbinden, H.},
  \bibinfo{author}{Legr{\'e}, M.}, and \bibinfo{author}{Gisin, N.}
  (\bibinfo{year}{2004}). \bibinfo{title}{Distribution of time-bin entangled
  qubits over 50 km of optical fiber}.
\newblock \bibinfo{journal}{Physical review letters}
  \emph{\bibinfo{volume}{93}}, \bibinfo{pages}{180502}.
\bibitem[{Takesue and Inoue(2005)}]{takesue2005generation}
\bibinfo{author}{Takesue, H.}, and \bibinfo{author}{Inoue, K.}
  (\bibinfo{year}{2005}). \bibinfo{title}{Generation of 1.5-$\mu$ m band
  time-bin entanglement using spontaneous fiber four-wave mixing and planar
  light-wave circuit interferometers}.
\newblock \bibinfo{journal}{Physical Review A—Atomic, Molecular, and Optical
  Physics} \emph{\bibinfo{volume}{72}}, \bibinfo{pages}{041804}.
\bibitem[{Honjo et~al.(2007)Honjo, Takesue, Kamada, Nishida, Tadanaga, Asobe
  and Inoue}]{honjo2007long}
\bibinfo{author}{Honjo, T.}, \bibinfo{author}{Takesue, H.},
  \bibinfo{author}{Kamada, H.}, \bibinfo{author}{Nishida, Y.},
  \bibinfo{author}{Tadanaga, O.}, \bibinfo{author}{Asobe, M.}, and
  \bibinfo{author}{Inoue, K.} (\bibinfo{year}{2007}).
  \bibinfo{title}{Long-distance distribution of time-bin entangled photon pairs
  over 100 km using frequency up-conversion detectors}.
\newblock \bibinfo{journal}{Optics express} \emph{\bibinfo{volume}{15}},
  \bibinfo{pages}{13957--13964}.
\bibitem[{Dynes et~al.(2009)Dynes, Takesue, Yuan, Sharpe, Harada, Honjo,
  Kamada, Tadanaga, Nishida, Asobe et~al.}]{dynes2009efficient}
\bibinfo{author}{Dynes, J.F.}, \bibinfo{author}{Takesue, H.},
  \bibinfo{author}{Yuan, Z.L.}, \bibinfo{author}{Sharpe, A.W.},
  \bibinfo{author}{Harada, K.}, \bibinfo{author}{Honjo, T.},
  \bibinfo{author}{Kamada, H.}, \bibinfo{author}{Tadanaga, O.},
  \bibinfo{author}{Nishida, Y.}, \bibinfo{author}{Asobe, M.} et~al.
  (\bibinfo{year}{2009}). \bibinfo{title}{Efficient entanglement distribution
  over 200 kilometers}.
\newblock \bibinfo{journal}{Optics express} \emph{\bibinfo{volume}{17}},
  \bibinfo{pages}{11440--11449}.
\bibitem[{Inagaki et~al.(2013)Inagaki, Matsuda, Tadanaga, Asobe and
  Takesue}]{inagaki2013entanglement}
\bibinfo{author}{Inagaki, T.}, \bibinfo{author}{Matsuda, N.},
  \bibinfo{author}{Tadanaga, O.}, \bibinfo{author}{Asobe, M.}, and
  \bibinfo{author}{Takesue, H.} (\bibinfo{year}{2013}).
  \bibinfo{title}{Entanglement distribution over 300 km of fiber}.
\newblock \bibinfo{journal}{Optics express} \emph{\bibinfo{volume}{21}},
  \bibinfo{pages}{23241--23249}.
\bibitem[{Shi et~al.(2023)Shi, Gerrits and Slattery}]{shi2023towards}
\bibinfo{author}{Shi, Y.}, \bibinfo{author}{Gerrits, T.}, and
  \bibinfo{author}{Slattery, O.} (\bibinfo{year}{2023}).
  \bibinfo{title}{Towards continuous fiber birefringence compensation with
  single-photon-level light}.
\newblock In \bibinfo{booktitle}{CLEO: Science and Innovations}.
  \bibinfo{organization}{Optica Publishing Group} pp.
  \bibinfo{pages}{JTh2A--24}.
\bibitem[{Jin et~al.(2010)Jin, Ren, Yang, Yi, Zhou, Xu, Wang, Yang, Hu, Jiang
  et~al.}]{jin2010experimental}
\bibinfo{author}{Jin, X.M.}, \bibinfo{author}{Ren, J.G.},
  \bibinfo{author}{Yang, B.}, \bibinfo{author}{Yi, Z.H.},
  \bibinfo{author}{Zhou, F.}, \bibinfo{author}{Xu, X.F.},
  \bibinfo{author}{Wang, S.K.}, \bibinfo{author}{Yang, D.},
  \bibinfo{author}{Hu, Y.F.}, \bibinfo{author}{Jiang, S.} et~al.
  (\bibinfo{year}{2010}). \bibinfo{title}{Experimental free-space quantum
  teleportation}.
\newblock \bibinfo{journal}{Nature photonics} \emph{\bibinfo{volume}{4}},
  \bibinfo{pages}{376--381}.
\bibitem[{Ursin et~al.(2006)Ursin, Tiefenbacher, Schmitt-Manderbach, Weier,
  Scheidl, Lindenthal, Blauensteiner, Jennewein, Perdigues, Trojek
  et~al.}]{ursin2006free}
\bibinfo{author}{Ursin, R.}, \bibinfo{author}{Tiefenbacher, F.},
  \bibinfo{author}{Schmitt-Manderbach, T.}, \bibinfo{author}{Weier, H.},
  \bibinfo{author}{Scheidl, T.}, \bibinfo{author}{Lindenthal, M.},
  \bibinfo{author}{Blauensteiner, B.}, \bibinfo{author}{Jennewein, T.},
  \bibinfo{author}{Perdigues, J.}, \bibinfo{author}{Trojek, P.} et~al.
  (\bibinfo{year}{2006}). \bibinfo{title}{Free-space distribution of
  entanglement and single photons over 144 km}.
\newblock \bibinfo{journal}{arXiv preprint quant-ph/0607182}.
\bibitem[{Ma et~al.(2012)Ma, Herbst, Scheidl, Wang, Kropatschek, Naylor,
  Wittmann, Mech, Kofler, Anisimova et~al.}]{ma2012quantum}
\bibinfo{author}{Ma, X.S.}, \bibinfo{author}{Herbst, T.},
  \bibinfo{author}{Scheidl, T.}, \bibinfo{author}{Wang, D.},
  \bibinfo{author}{Kropatschek, S.}, \bibinfo{author}{Naylor, W.},
  \bibinfo{author}{Wittmann, B.}, \bibinfo{author}{Mech, A.},
  \bibinfo{author}{Kofler, J.}, \bibinfo{author}{Anisimova, E.} et~al.
  (\bibinfo{year}{2012}). \bibinfo{title}{Quantum teleportation over 143
  kilometres using active feed-forward}.
\newblock \bibinfo{journal}{Nature} \emph{\bibinfo{volume}{489}},
  \bibinfo{pages}{269--273}.
\bibitem[{Xingjian et~al.(2025)Xingjian, Haoran, Chua, Eng, Meunier, Grieve,
  Weibo and Ling}]{xingjian2025polarization}
\bibinfo{author}{Xingjian, Z.}, \bibinfo{author}{Haoran, Z.},
  \bibinfo{author}{Chua, R.M.}, \bibinfo{author}{Eng, J.},
  \bibinfo{author}{Meunier, M.}, \bibinfo{author}{Grieve, J.A.},
  \bibinfo{author}{Weibo, G.}, and \bibinfo{author}{Ling, A.}
  (\bibinfo{year}{2025}). \bibinfo{title}{Polarization-encoded quantum key
  distribution with a room-temperature telecom single-photon emitter}.
\newblock \bibinfo{journal}{National Science Review} pp.
  \bibinfo{pages}{nwaf147}.
\bibitem[{Rodimin et~al.(2024)Rodimin, Kravtsov, Chua, De~Santis, Ponasenko,
  Kurochkin, Ling and Grieve}]{rodimin2024impact}
\bibinfo{author}{Rodimin, V.}, \bibinfo{author}{Kravtsov, K.},
  \bibinfo{author}{Chua, R.M.}, \bibinfo{author}{De~Santis, G.},
  \bibinfo{author}{Ponasenko, A.}, \bibinfo{author}{Kurochkin, Y.},
  \bibinfo{author}{Ling, A.}, and \bibinfo{author}{Grieve, J.A.}
  (\bibinfo{year}{2024}). \bibinfo{title}{Impact of polarization mode
  dispersion on entangled photon distribution}.
\newblock \bibinfo{journal}{arXiv preprint arXiv:2408.01754}.
\bibitem[{Wengerowsky et~al.(2019)Wengerowsky, Joshi, Steinlechner, Zichi,
  Dobrovolskiy, Van~der Molen, Los, Zwiller, Versteegh, Mura
  et~al.}]{wengerowsky2019entanglement}
\bibinfo{author}{Wengerowsky, S.}, \bibinfo{author}{Joshi, S.K.},
  \bibinfo{author}{Steinlechner, F.}, \bibinfo{author}{Zichi, J.R.},
  \bibinfo{author}{Dobrovolskiy, S.M.}, \bibinfo{author}{Van~der Molen, R.},
  \bibinfo{author}{Los, J.W.}, \bibinfo{author}{Zwiller, V.},
  \bibinfo{author}{Versteegh, M.A.}, \bibinfo{author}{Mura, A.} et~al.
  (\bibinfo{year}{2019}). \bibinfo{title}{Entanglement distribution over a
  96-km-long submarine optical fiber}.
\newblock \bibinfo{journal}{Proceedings of the National Academy of Sciences}
  \emph{\bibinfo{volume}{116}}, \bibinfo{pages}{6684--6688}.
\bibitem[{Wengerowsky et~al.(2020)Wengerowsky, Joshi, Steinlechner, Zichi, Liu,
  Scheidl, Dobrovolskiy, Molen, Los, Zwiller et~al.}]{wengerowsky2020passively}
\bibinfo{author}{Wengerowsky, S.}, \bibinfo{author}{Joshi, S.K.},
  \bibinfo{author}{Steinlechner, F.}, \bibinfo{author}{Zichi, J.R.},
  \bibinfo{author}{Liu, B.}, \bibinfo{author}{Scheidl, T.},
  \bibinfo{author}{Dobrovolskiy, S.M.}, \bibinfo{author}{Molen, R.v.d.},
  \bibinfo{author}{Los, J.W.}, \bibinfo{author}{Zwiller, V.} et~al.
  (\bibinfo{year}{2020}). \bibinfo{title}{Passively stable distribution of
  polarisation entanglement over 192 km of deployed optical fibre}.
\newblock \bibinfo{journal}{npj Quantum Information}
  \emph{\bibinfo{volume}{6}}, \bibinfo{pages}{5}.
\bibitem[{Neumann et~al.(2022)Neumann, Buchner, Bulla, Bohmann and
  Ursin}]{neumann2022continuous}
\bibinfo{author}{Neumann, S.P.}, \bibinfo{author}{Buchner, A.},
  \bibinfo{author}{Bulla, L.}, \bibinfo{author}{Bohmann, M.}, and
  \bibinfo{author}{Ursin, R.} (\bibinfo{year}{2022}).
  \bibinfo{title}{Continuous entanglement distribution over a transnational 248
  km fiber link}.
\newblock \bibinfo{journal}{Nature Communications} \emph{\bibinfo{volume}{13}},
  \bibinfo{pages}{6134}.
\bibitem[{Rahmouni et~al.(2024{\natexlab{a}})Rahmouni, Kuo, Li-Baboud,
  Burenkov, Shi, Jabir, Lal, Reddy, Merzouki, Ma
  et~al.}]{rahmouni2024metropolitan}
\bibinfo{author}{Rahmouni, A.}, \bibinfo{author}{Kuo, P.},
  \bibinfo{author}{Li-Baboud, Y.S.}, \bibinfo{author}{Burenkov, I.},
  \bibinfo{author}{Shi, Y.}, \bibinfo{author}{Jabir, M.}, \bibinfo{author}{Lal,
  N.}, \bibinfo{author}{Reddy, D.}, \bibinfo{author}{Merzouki, M.},
  \bibinfo{author}{Ma, L.} et~al. (\bibinfo{year}{2024}{\natexlab{a}}).
  \bibinfo{title}{Metropolitan-scale entanglement distribution with co existing
  quantum and classical signals in a single fiber}.
\newblock \bibinfo{journal}{arXiv preprint arXiv:2402.00617}.
\bibitem[{Zhuang et~al.(2024)Zhuang, Li, Zheng, Zeng, Wu, Li, Yao, Xie, Li, Qin
  et~al.}]{zhuang2024ultrabright}
\bibinfo{author}{Zhuang, S.C.}, \bibinfo{author}{Li, B.},
  \bibinfo{author}{Zheng, M.Y.}, \bibinfo{author}{Zeng, Y.X.},
  \bibinfo{author}{Wu, H.N.}, \bibinfo{author}{Li, G.B.}, \bibinfo{author}{Yao,
  Q.}, \bibinfo{author}{Xie, X.P.}, \bibinfo{author}{Li, Y.H.},
  \bibinfo{author}{Qin, H.} et~al. (\bibinfo{year}{2024}).
  \bibinfo{title}{Ultrabright-entanglement-based quantum key distribution over
  a 404-km-long optical}.
\newblock \bibinfo{journal}{arXiv preprint arXiv:2408.04361}.
\bibitem[{Llewellyn et~al.(2020)Llewellyn, Ding, Faruque, Paesani, Bacco,
  Santagati, Qian, Li, Xiao, Huber et~al.}]{llewellyn2020chip}
\bibinfo{author}{Llewellyn, D.}, \bibinfo{author}{Ding, Y.},
  \bibinfo{author}{Faruque, I.I.}, \bibinfo{author}{Paesani, S.},
  \bibinfo{author}{Bacco, D.}, \bibinfo{author}{Santagati, R.},
  \bibinfo{author}{Qian, Y.J.}, \bibinfo{author}{Li, Y.},
  \bibinfo{author}{Xiao, Y.F.}, \bibinfo{author}{Huber, M.} et~al.
  (\bibinfo{year}{2020}). \bibinfo{title}{Chip-to-chip quantum teleportation
  and multi-photon entanglement in silicon}.
\newblock \bibinfo{journal}{Nature Physics} \emph{\bibinfo{volume}{16}},
  \bibinfo{pages}{148--153}.
\bibitem[{Zheng et~al.(2023)Zheng, Zhai, Liu, Mao, Chen, Dai, Huang, Bao, Fu,
  Tong et~al.}]{zheng2023multichip}
\bibinfo{author}{Zheng, Y.}, \bibinfo{author}{Zhai, C.}, \bibinfo{author}{Liu,
  D.}, \bibinfo{author}{Mao, J.}, \bibinfo{author}{Chen, X.},
  \bibinfo{author}{Dai, T.}, \bibinfo{author}{Huang, J.}, \bibinfo{author}{Bao,
  J.}, \bibinfo{author}{Fu, Z.}, \bibinfo{author}{Tong, Y.} et~al.
  (\bibinfo{year}{2023}). \bibinfo{title}{Multichip multidimensional quantum
  networks with entanglement retrievability}.
\newblock \bibinfo{journal}{Science} \emph{\bibinfo{volume}{381}},
  \bibinfo{pages}{221--226}.
\bibitem[{Du et~al.(2024)Du, Chen, Gao, Grieve, Tan and
  Ling}]{du2024demonstration}
\bibinfo{author}{Du, J.}, \bibinfo{author}{Chen, G.F.}, \bibinfo{author}{Gao,
  H.}, \bibinfo{author}{Grieve, J.A.}, \bibinfo{author}{Tan, D.T.}, and
  \bibinfo{author}{Ling, A.} (\bibinfo{year}{2024}).
  \bibinfo{title}{Demonstration of a low loss, highly stable and re-useable
  edge coupler for high heralding efficiency and low g (2)(0) soi correlated
  photon pair sources}.
\newblock \bibinfo{journal}{Optics Express} \emph{\bibinfo{volume}{32}},
  \bibinfo{pages}{11406--11418}.
\bibitem[{Suo et~al.(2015)Suo, Dong, Zhang, Huang and Peng}]{suo2015generation}
\bibinfo{author}{Suo, J.}, \bibinfo{author}{Dong, S.}, \bibinfo{author}{Zhang,
  W.}, \bibinfo{author}{Huang, Y.}, and \bibinfo{author}{Peng, J.}
  (\bibinfo{year}{2015}). \bibinfo{title}{Generation of hyper-entanglement on
  polarization and energy-time based on a silicon micro-ring cavity}.
\newblock \bibinfo{journal}{Optics express} \emph{\bibinfo{volume}{23}},
  \bibinfo{pages}{3985--3995}.
\bibitem[{Takesue et~al.(2008)Takesue, Fukuda, Tsuchizawa, Watanabe, Yamada,
  Tokura and Itabashi}]{takesue2008generation}
\bibinfo{author}{Takesue, H.}, \bibinfo{author}{Fukuda, H.},
  \bibinfo{author}{Tsuchizawa, T.}, \bibinfo{author}{Watanabe, T.},
  \bibinfo{author}{Yamada, K.}, \bibinfo{author}{Tokura, Y.}, and
  \bibinfo{author}{Itabashi, S.i.} (\bibinfo{year}{2008}).
  \bibinfo{title}{Generation of polarization entangled photon pairs using
  silicon wire waveguide}.
\newblock \bibinfo{journal}{Optics express} \emph{\bibinfo{volume}{16}},
  \bibinfo{pages}{5721--5727}.
\bibitem[{Li et~al.(2017)Li, Zhou, Feng, Fang, Liu, Liu, Wang, Ren, Ding, Xu
  et~al.}]{li2017chip}
\bibinfo{author}{Li, Y.H.}, \bibinfo{author}{Zhou, Z.Y.},
  \bibinfo{author}{Feng, L.T.}, \bibinfo{author}{Fang, W.T.},
  \bibinfo{author}{Liu, S.l.}, \bibinfo{author}{Liu, S.K.},
  \bibinfo{author}{Wang, K.}, \bibinfo{author}{Ren, X.F.},
  \bibinfo{author}{Ding, D.S.}, \bibinfo{author}{Xu, L.X.} et~al.
  (\bibinfo{year}{2017}). \bibinfo{title}{On-chip multiplexed multiple
  entanglement sources in a single silicon nanowire}.
\newblock \bibinfo{journal}{Physical Review Applied}
  \emph{\bibinfo{volume}{7}}, \bibinfo{pages}{064005}.
\bibitem[{Sharma et~al.(2022)Sharma, Venkataraman and
  Ghosh}]{sharma2022silicon}
\bibinfo{author}{Sharma, S.}, \bibinfo{author}{Venkataraman, V.}, and
  \bibinfo{author}{Ghosh, J.} (\bibinfo{year}{2022}). \bibinfo{title}{Silicon
  photonic wires for broadband polarization entanglement at telecommunication
  wavelengths}.
\newblock \bibinfo{journal}{Physical Review Applied}
  \emph{\bibinfo{volume}{18}}, \bibinfo{pages}{044043}.
\bibitem[{Miloshevsky et~al.(2024)Miloshevsky, Cohen, Myilswamy, Alshowkan,
  Fatema, Lu, Weiner and Lukens}]{miloshevsky2024cmos}
\bibinfo{author}{Miloshevsky, A.}, \bibinfo{author}{Cohen, L.M.},
  \bibinfo{author}{Myilswamy, K.V.}, \bibinfo{author}{Alshowkan, M.},
  \bibinfo{author}{Fatema, S.}, \bibinfo{author}{Lu, H.H.},
  \bibinfo{author}{Weiner, A.M.}, and \bibinfo{author}{Lukens, J.M.}
  (\bibinfo{year}{2024}). \bibinfo{title}{Cmos photonic integrated source of
  broadband polarization-entangled photons}.
\newblock \bibinfo{journal}{Optica Quantum} \emph{\bibinfo{volume}{2}},
  \bibinfo{pages}{254--259}.
\bibitem[{Wen et~al.(2023)Wen, Yan, Lu, Lu, Wu, Lu, Zhu and
  Ma}]{wen2023polarization}
\bibinfo{author}{Wen, W.}, \bibinfo{author}{Yan, W.}, \bibinfo{author}{Lu, C.},
  \bibinfo{author}{Lu, L.}, \bibinfo{author}{Wu, X.}, \bibinfo{author}{Lu, Y.},
  \bibinfo{author}{Zhu, S.}, and \bibinfo{author}{Ma, X.S.}
  (\bibinfo{year}{2023}). \bibinfo{title}{Polarization-entangled quantum
  frequency comb from a silicon nitride microring resonator}.
\newblock \bibinfo{journal}{Physical Review Applied}
  \emph{\bibinfo{volume}{20}}, \bibinfo{pages}{064032}.
\bibitem[{Zhang et~al.(2019)Zhang, Feng, Zhou, Chen, Wu, Li, Gao, Guo, Guo, Dai
  et~al.}]{zhang2019generation}
\bibinfo{author}{Zhang, M.}, \bibinfo{author}{Feng, L.T.},
  \bibinfo{author}{Zhou, Z.Y.}, \bibinfo{author}{Chen, Y.},
  \bibinfo{author}{Wu, H.}, \bibinfo{author}{Li, M.}, \bibinfo{author}{Gao,
  S.M.}, \bibinfo{author}{Guo, G.P.}, \bibinfo{author}{Guo, G.C.},
  \bibinfo{author}{Dai, D.X.} et~al. (\bibinfo{year}{2019}).
  \bibinfo{title}{Generation of multiphoton quantum states on silicon}.
\newblock \bibinfo{journal}{Light: Science \& Applications}
  \emph{\bibinfo{volume}{8}}, \bibinfo{pages}{41}.
\bibitem[{Chua et~al.(2022)Chua, Grieve and Ling}]{chua2022fine}
\bibinfo{author}{Chua, R.}, \bibinfo{author}{Grieve, J.A.}, and
  \bibinfo{author}{Ling, A.} (\bibinfo{year}{2022}).
  \bibinfo{title}{Fine-grained all-fiber nonlocal dispersion compensation in
  the telecommunications o-band}.
\newblock \bibinfo{journal}{Optics Express} \emph{\bibinfo{volume}{30}},
  \bibinfo{pages}{15607--15615}.
\bibitem[{Neumann et~al.(2021)Neumann, Ribezzo, Bohmann and
  Ursin}]{neumann2021experimentally}
\bibinfo{author}{Neumann, S.P.}, \bibinfo{author}{Ribezzo, D.},
  \bibinfo{author}{Bohmann, M.}, and \bibinfo{author}{Ursin, R.}
  (\bibinfo{year}{2021}). \bibinfo{title}{Experimentally optimizing qkd rates
  via nonlocal dispersion compensation}.
\newblock \bibinfo{journal}{Quantum Science and Technology}
  \emph{\bibinfo{volume}{6}}, \bibinfo{pages}{025017}.
\bibitem[{Boyd et~al.(2008)Boyd, Gaeta and Giese}]{boyd2008nonlinear}
\bibinfo{author}{Boyd, R.W.}, \bibinfo{author}{Gaeta, A.L.}, and
  \bibinfo{author}{Giese, E.} (\bibinfo{year}{2008}). \bibinfo{title}{Nonlinear
  optics}.
\newblock In \bibinfo{booktitle}{Springer Handbook of Atomic, Molecular, and
  Optical Physics} pp. \bibinfo{pages}{1097--1110}..
  \bibinfo{publisher}{Springer} pp. \bibinfo{pages}{1097--1110}.
\bibitem[{Appas et~al.(2021)Appas, Baboux, Amanti, Lema{\'\i}tre, Boitier,
  Diamanti and Ducci}]{appas2021flexible}
\bibinfo{author}{Appas, F.}, \bibinfo{author}{Baboux, F.},
  \bibinfo{author}{Amanti, M.I.}, \bibinfo{author}{Lema{\'\i}tre, A.},
  \bibinfo{author}{Boitier, F.}, \bibinfo{author}{Diamanti, E.}, and
  \bibinfo{author}{Ducci, S.} (\bibinfo{year}{2021}). \bibinfo{title}{Flexible
  entanglement-distribution network with an algaas chip for secure
  communications}.
\newblock \bibinfo{journal}{npj Quantum Information}
  \emph{\bibinfo{volume}{7}}, \bibinfo{pages}{118}.
\bibitem[{Meyer-Scott et~al.(2018)Meyer-Scott, Prasannan, Eigner, Quiring,
  Donohue, Barkhofen and Silberhorn}]{meyer2018high}
\bibinfo{author}{Meyer-Scott, E.}, \bibinfo{author}{Prasannan, N.},
  \bibinfo{author}{Eigner, C.}, \bibinfo{author}{Quiring, V.},
  \bibinfo{author}{Donohue, J.M.}, \bibinfo{author}{Barkhofen, S.}, and
  \bibinfo{author}{Silberhorn, C.} (\bibinfo{year}{2018}).
  \bibinfo{title}{High-performance source of spectrally pure, polarization
  entangled photon pairs based on hybrid integrated-bulk optics}.
\newblock \bibinfo{journal}{Optics express} \emph{\bibinfo{volume}{26}},
  \bibinfo{pages}{32475--32490}.
\bibitem[{Gisin et~al.(2002)Gisin, Ribordy, Tittel and
  Zbinden}]{gisin2002quantum}
\bibinfo{author}{Gisin, N.}, \bibinfo{author}{Ribordy, G.},
  \bibinfo{author}{Tittel, W.}, and \bibinfo{author}{Zbinden, H.}
  (\bibinfo{year}{2002}). \bibinfo{title}{Quantum cryptography}.
\newblock \bibinfo{journal}{Reviews of modern physics}
  \emph{\bibinfo{volume}{74}}, \bibinfo{pages}{145}.
\bibitem[{Shi et~al.(2020)Shi, Moe~Thar, Poh, Grieve, Kurtsiefer and
  Ling}]{shi2020stable}
\bibinfo{author}{Shi, Y.}, \bibinfo{author}{Moe~Thar, S.},
  \bibinfo{author}{Poh, H.S.}, \bibinfo{author}{Grieve, J.A.},
  \bibinfo{author}{Kurtsiefer, C.}, and \bibinfo{author}{Ling, A.}
  (\bibinfo{year}{2020}). \bibinfo{title}{Stable polarization entanglement
  based quantum key distribution over a deployed metropolitan fiber}.
\newblock \bibinfo{journal}{Applied Physics Letters}
  \emph{\bibinfo{volume}{117}}.
\bibitem[{Zhang et~al.(2024{\natexlab{a}})Zhang, Zhang, Chua, Eng, Meunier,
  Grieve, Gao and Ling}]{zhang2024polarizationencodedquantumkeydistribution}
\bibinfo{author}{Zhang, X.}, \bibinfo{author}{Zhang, H.},
  \bibinfo{author}{Chua, R.M.}, \bibinfo{author}{Eng, J.},
  \bibinfo{author}{Meunier, M.}, \bibinfo{author}{Grieve, J.A.},
  \bibinfo{author}{Gao, W.}, and \bibinfo{author}{Ling, A.}
  (\bibinfo{year}{2024}{\natexlab{a}}).
\newblock \bibinfo{title}{Polarization-encoded quantum key distribution with a
  room-temperature telecom single-photon emitter}. {\natexlab{a}}.
\newblock \URLprefix \url{https://arxiv.org/abs/2409.17060}.
  \href{http://arxiv.org/abs/2409.17060}{\tt arXiv:2409.17060}.
\bibitem[{Antonelli et~al.(2011)Antonelli, Shtaif and
  Brodsky}]{antonelli2011sudden}
\bibinfo{author}{Antonelli, C.}, \bibinfo{author}{Shtaif, M.}, and
  \bibinfo{author}{Brodsky, M.} (\bibinfo{year}{2011}). \bibinfo{title}{Sudden
  death of entanglement induced by polarization mode dispersion}.
\newblock \bibinfo{journal}{Physical review letters}
  \emph{\bibinfo{volume}{106}}, \bibinfo{pages}{080404}.
\bibitem[{Steiner et~al.(2023)Steiner, Shen, Castro, Bowers and
  Moody}]{steiner2023continuous}
\bibinfo{author}{Steiner, T.J.}, \bibinfo{author}{Shen, M.},
  \bibinfo{author}{Castro, J.E.}, \bibinfo{author}{Bowers, J.E.}, and
  \bibinfo{author}{Moody, G.} (\bibinfo{year}{2023}).
  \bibinfo{title}{Continuous entanglement distribution from an
  algaas-on-insulator microcomb for quantum communications}.
\newblock \bibinfo{journal}{Optica Quantum} \emph{\bibinfo{volume}{1}},
  \bibinfo{pages}{55--62}.
\bibitem[{Autebert et~al.(2016)Autebert, Bruno, Martin, Lemaitre, Carbonell,
  Favero, Leo, Zbinden and Ducci}]{autebert2016integrated}
\bibinfo{author}{Autebert, C.}, \bibinfo{author}{Bruno, N.},
  \bibinfo{author}{Martin, A.}, \bibinfo{author}{Lemaitre, A.},
  \bibinfo{author}{Carbonell, C.G.}, \bibinfo{author}{Favero, I.},
  \bibinfo{author}{Leo, G.}, \bibinfo{author}{Zbinden, H.}, and
  \bibinfo{author}{Ducci, S.} (\bibinfo{year}{2016}).
  \bibinfo{title}{Integrated algaas source of highly indistinguishable and
  energy-time entangled photons}.
\newblock \bibinfo{journal}{Optica} \emph{\bibinfo{volume}{3}},
  \bibinfo{pages}{143--146}.
\bibitem[{Steiner et~al.(2021)Steiner, Castro, Chang, Dang, Xie, Norman, Bowers
  and Moody}]{steiner2021ultrabright}
\bibinfo{author}{Steiner, T.J.}, \bibinfo{author}{Castro, J.E.},
  \bibinfo{author}{Chang, L.}, \bibinfo{author}{Dang, Q.},
  \bibinfo{author}{Xie, W.}, \bibinfo{author}{Norman, J.},
  \bibinfo{author}{Bowers, J.E.}, and \bibinfo{author}{Moody, G.}
  (\bibinfo{year}{2021}). \bibinfo{title}{Ultrabright entangled-photon-pair
  generation from an al ga as-on-insulator microring resonator}.
\newblock \bibinfo{journal}{Prx Quantum} \emph{\bibinfo{volume}{2}},
  \bibinfo{pages}{010337}.
\bibitem[{Shi et~al.(2024)Shi, Mohanraj, Dhyani, Baiju, Wang, Sun, Zhou,
  Paterova, Leong and Zhu}]{shi2024efficient}
\bibinfo{author}{Shi, X.}, \bibinfo{author}{Mohanraj, S.S.},
  \bibinfo{author}{Dhyani, V.}, \bibinfo{author}{Baiju, A.A.},
  \bibinfo{author}{Wang, S.}, \bibinfo{author}{Sun, J.}, \bibinfo{author}{Zhou,
  L.}, \bibinfo{author}{Paterova, A.}, \bibinfo{author}{Leong, V.}, and
  \bibinfo{author}{Zhu, D.} (\bibinfo{year}{2024}). \bibinfo{title}{Efficient
  photon-pair generation in layer-poled lithium niobate nanophotonic
  waveguides}.
\newblock \bibinfo{journal}{Light: Science \& Applications}
  \emph{\bibinfo{volume}{13}}, \bibinfo{pages}{282}.
\bibitem[{Zhao et~al.(2020)Zhao, Ma, R{\"u}sing and Mookherjea}]{zhao2020high}
\bibinfo{author}{Zhao, J.}, \bibinfo{author}{Ma, C.},
  \bibinfo{author}{R{\"u}sing, M.}, and \bibinfo{author}{Mookherjea, S.}
  (\bibinfo{year}{2020}). \bibinfo{title}{High quality entangled photon pair
  generation in periodically poled thin-film lithium niobate waveguides}.
\newblock \bibinfo{journal}{Physical review letters}
  \emph{\bibinfo{volume}{124}}, \bibinfo{pages}{163603}.
\bibitem[{Lohrmann et~al.(2017)Lohrmann, Johnson, McCallum and
  Castelletto}]{lohrmann2017review}
\bibinfo{author}{Lohrmann, A.}, \bibinfo{author}{Johnson, B.},
  \bibinfo{author}{McCallum, J.}, and \bibinfo{author}{Castelletto, S.}
  (\bibinfo{year}{2017}). \bibinfo{title}{A review on single photon sources in
  silicon carbide}.
\newblock \bibinfo{journal}{Reports on Progress in Physics}
  \emph{\bibinfo{volume}{80}}, \bibinfo{pages}{034502}.
\bibitem[{Rahmouni et~al.(2024{\natexlab{b}})Rahmouni, Wang, Li, Tang, Gerrits,
  Slattery, Li and Ma}]{rahmouni2024entangled}
\bibinfo{author}{Rahmouni, A.}, \bibinfo{author}{Wang, R.},
  \bibinfo{author}{Li, J.}, \bibinfo{author}{Tang, X.},
  \bibinfo{author}{Gerrits, T.}, \bibinfo{author}{Slattery, O.},
  \bibinfo{author}{Li, Q.}, and \bibinfo{author}{Ma, L.}
  (\bibinfo{year}{2024}{\natexlab{b}}). \bibinfo{title}{Entangled photon pair
  generation in an integrated sic platform}.
\newblock \bibinfo{journal}{Light: Science \& Applications}
  \emph{\bibinfo{volume}{13}}, \bibinfo{pages}{110}.
\bibitem[{Choi et~al.(2020)Choi, Sohn, Chen, Ng and Tan}]{choi2020correlated}
\bibinfo{author}{Choi, J.W.}, \bibinfo{author}{Sohn, B.U.},
  \bibinfo{author}{Chen, G.F.}, \bibinfo{author}{Ng, D.K.}, and
  \bibinfo{author}{Tan, D.T.} (\bibinfo{year}{2020}).
  \bibinfo{title}{Correlated photon pair generation in ultra-silicon-rich
  nitride waveguide}.
\newblock \bibinfo{journal}{Optics Communications}
  \emph{\bibinfo{volume}{463}}, \bibinfo{pages}{125351}.
\bibitem[{Guo et~al.(2017)Guo, Zou, Schuck, Jung, Cheng and
  Tang}]{guo2017parametric}
\bibinfo{author}{Guo, X.}, \bibinfo{author}{Zou, C.l.},
  \bibinfo{author}{Schuck, C.}, \bibinfo{author}{Jung, H.},
  \bibinfo{author}{Cheng, R.}, and \bibinfo{author}{Tang, H.X.}
  (\bibinfo{year}{2017}). \bibinfo{title}{Parametric down-conversion
  photon-pair source on a nanophotonic chip}.
\newblock \bibinfo{journal}{Light: Science \& Applications}
  \emph{\bibinfo{volume}{6}}, \bibinfo{pages}{e16249--e16249}.
\bibitem[{Lu et~al.(2022)Lu, Cao, Peng and Pan}]{lu2022micius}
\bibinfo{author}{Lu, C.Y.}, \bibinfo{author}{Cao, Y.}, \bibinfo{author}{Peng,
  C.Z.}, and \bibinfo{author}{Pan, J.W.} (\bibinfo{year}{2022}).
  \bibinfo{title}{Micius quantum experiments in space}.
\newblock \bibinfo{journal}{Reviews of Modern Physics}
  \emph{\bibinfo{volume}{94}}, \bibinfo{pages}{035001}.
\bibitem[{Ong et~al.(2013)Ong, Kumar and Mookherjea}]{ong2013ultra}
\bibinfo{author}{Ong, J.R.}, \bibinfo{author}{Kumar, R.}, and
  \bibinfo{author}{Mookherjea, S.} (\bibinfo{year}{2013}).
  \bibinfo{title}{Ultra-high-contrast and tunable-bandwidth filter using
  cascaded high-order silicon microring filters}.
\newblock \bibinfo{journal}{IEEE Photonics Technology Letters}
  \emph{\bibinfo{volume}{25}}, \bibinfo{pages}{1543--1546}.
\bibitem[{Luo et~al.(2014)Luo, Ophir, Chen, Gabrielli, Poitras, Bergmen and
  Lipson}]{luo2014wdm}
\bibinfo{author}{Luo, L.W.}, \bibinfo{author}{Ophir, N.},
  \bibinfo{author}{Chen, C.P.}, \bibinfo{author}{Gabrielli, L.H.},
  \bibinfo{author}{Poitras, C.B.}, \bibinfo{author}{Bergmen, K.}, and
  \bibinfo{author}{Lipson, M.} (\bibinfo{year}{2014}).
  \bibinfo{title}{Wdm-compatible mode-division multiplexing on a silicon chip}.
\newblock \bibinfo{journal}{Nature communications} \emph{\bibinfo{volume}{5}},
  \bibinfo{pages}{1--7}.
\bibitem[{Kumar et~al.(2020)Kumar, Wu and Tsang}]{kumar2020compact}
\bibinfo{author}{Kumar, R.R.}, \bibinfo{author}{Wu, X.}, and
  \bibinfo{author}{Tsang, H.K.} (\bibinfo{year}{2020}). \bibinfo{title}{Compact
  high-extinction tunable crow filters for integrated quantum photonic
  circuits}.
\newblock \bibinfo{journal}{Optics Letters} \emph{\bibinfo{volume}{45}},
  \bibinfo{pages}{1289--1292}.
\bibitem[{Cui et~al.(2022)Cui, Wang, Li, Ma, Ou and Li}]{cui2022programmable}
\bibinfo{author}{Cui, L.}, \bibinfo{author}{Wang, J.}, \bibinfo{author}{Li,
  J.}, \bibinfo{author}{Ma, M.}, \bibinfo{author}{Ou, Z.}, and
  \bibinfo{author}{Li, X.} (\bibinfo{year}{2022}). \bibinfo{title}{Programmable
  photon pair source}.
\newblock \bibinfo{journal}{APL Photonics} \emph{\bibinfo{volume}{7}}.
\bibitem[{Ong et~al.(2022)Ong, Chen, Xing, Gao and Tan}]{ong2022dispersion}
\bibinfo{author}{Ong, K.Y.K.}, \bibinfo{author}{Chen, G.F.},
  \bibinfo{author}{Xing, P.}, \bibinfo{author}{Gao, H.}, and
  \bibinfo{author}{Tan, D.T.} (\bibinfo{year}{2022}).
  \bibinfo{title}{Dispersion compensation of high-speed data using an
  integrated silicon nitride ring resonator}.
\newblock \bibinfo{journal}{Optics Express} \emph{\bibinfo{volume}{30}},
  \bibinfo{pages}{13959--13967}.
\bibitem[{Joshi et~al.(2018)Joshi, Farsi, Clemmen, Ramelow and
  Gaeta}]{joshi2018frequency}
\bibinfo{author}{Joshi, C.}, \bibinfo{author}{Farsi, A.},
  \bibinfo{author}{Clemmen, S.}, \bibinfo{author}{Ramelow, S.}, and
  \bibinfo{author}{Gaeta, A.L.} (\bibinfo{year}{2018}).
  \bibinfo{title}{Frequency multiplexing for quasi-deterministic heralded
  single-photon sources}.
\newblock \bibinfo{journal}{Nature communications} \emph{\bibinfo{volume}{9}},
  \bibinfo{pages}{847}.
\bibitem[{Xiong et~al.(2015)Xiong, Zhang, Mahendra, He, Choi, Chae, Marpaung,
  Leinse, Heideman, Hoekman et~al.}]{xiong2015compact}
\bibinfo{author}{Xiong, C.}, \bibinfo{author}{Zhang, X.},
  \bibinfo{author}{Mahendra, A.}, \bibinfo{author}{He, J.},
  \bibinfo{author}{Choi, D.Y.}, \bibinfo{author}{Chae, C.},
  \bibinfo{author}{Marpaung, D.}, \bibinfo{author}{Leinse, A.},
  \bibinfo{author}{Heideman, R.}, \bibinfo{author}{Hoekman, M.} et~al.
  (\bibinfo{year}{2015}). \bibinfo{title}{Compact and reconfigurable silicon
  nitride time-bin entanglement circuit}.
\newblock \bibinfo{journal}{Optica} \emph{\bibinfo{volume}{2}},
  \bibinfo{pages}{724--727}.
\bibitem[{Liu et~al.(2019)Liu, Yao, Wang, Li, Wang, You, Huang and
  Zhang}]{liu2019energy}
\bibinfo{author}{Liu, X.}, \bibinfo{author}{Yao, X.}, \bibinfo{author}{Wang,
  H.}, \bibinfo{author}{Li, H.}, \bibinfo{author}{Wang, Z.},
  \bibinfo{author}{You, L.}, \bibinfo{author}{Huang, Y.}, and
  \bibinfo{author}{Zhang, W.} (\bibinfo{year}{2019}).
  \bibinfo{title}{Energy-time entanglement-based dispersive optics quantum key
  distribution over optical fibers of 20 km}.
\newblock \bibinfo{journal}{Applied Physics Letters}
  \emph{\bibinfo{volume}{114}}.
\bibitem[{Liu et~al.(2023)Liu, Lin, Liu, Feng, Liu, Cui, Huang and
  Zhang}]{liu2023high}
\bibinfo{author}{Liu, J.}, \bibinfo{author}{Lin, Z.}, \bibinfo{author}{Liu,
  D.}, \bibinfo{author}{Feng, X.}, \bibinfo{author}{Liu, F.},
  \bibinfo{author}{Cui, K.}, \bibinfo{author}{Huang, Y.}, and
  \bibinfo{author}{Zhang, W.} (\bibinfo{year}{2023}).
  \bibinfo{title}{High-dimensional quantum key distribution using energy-time
  entanglement over 242 km partially deployed fiber}.
\newblock \bibinfo{journal}{Quantum Science and Technology}
  \emph{\bibinfo{volume}{9}}, \bibinfo{pages}{015003}.
\bibitem[{Wen et~al.(2022)Wen, Chen, Lu, Yan, Xue, Zhang, Lu, Zhu and
  Ma}]{wen2022realizing}
\bibinfo{author}{Wen, W.}, \bibinfo{author}{Chen, Z.}, \bibinfo{author}{Lu,
  L.}, \bibinfo{author}{Yan, W.}, \bibinfo{author}{Xue, W.},
  \bibinfo{author}{Zhang, P.}, \bibinfo{author}{Lu, Y.}, \bibinfo{author}{Zhu,
  S.}, and \bibinfo{author}{Ma, X.s.} (\bibinfo{year}{2022}).
  \bibinfo{title}{Realizing an entanglement-based multiuser quantum network
  with integrated photonics}.
\newblock \bibinfo{journal}{Physical Review Applied}
  \emph{\bibinfo{volume}{18}}, \bibinfo{pages}{024059}.
\bibitem[{Lu et~al.(2020)Lu, Xia, Chen, Chen, Yu, Tao, Ma, Pan, Cai, Lu
  et~al.}]{lu2020three}
\bibinfo{author}{Lu, L.}, \bibinfo{author}{Xia, L.}, \bibinfo{author}{Chen,
  Z.}, \bibinfo{author}{Chen, L.}, \bibinfo{author}{Yu, T.},
  \bibinfo{author}{Tao, T.}, \bibinfo{author}{Ma, W.}, \bibinfo{author}{Pan,
  Y.}, \bibinfo{author}{Cai, X.}, \bibinfo{author}{Lu, Y.} et~al.
  (\bibinfo{year}{2020}). \bibinfo{title}{Three-dimensional entanglement on a
  silicon chip}.
\newblock \bibinfo{journal}{npj Quantum Information}
  \emph{\bibinfo{volume}{6}}, \bibinfo{pages}{30}.
\bibitem[{Xie et~al.(2015)Xie, Zhong, Shrestha, Xu, Liang, Gong, Bienfang,
  Restelli, Shapiro, Wong et~al.}]{xie2015harnessing}
\bibinfo{author}{Xie, Z.}, \bibinfo{author}{Zhong, T.},
  \bibinfo{author}{Shrestha, S.}, \bibinfo{author}{Xu, X.},
  \bibinfo{author}{Liang, J.}, \bibinfo{author}{Gong, Y.X.},
  \bibinfo{author}{Bienfang, J.C.}, \bibinfo{author}{Restelli, A.},
  \bibinfo{author}{Shapiro, J.H.}, \bibinfo{author}{Wong, F.N.} et~al.
  (\bibinfo{year}{2015}). \bibinfo{title}{Harnessing high-dimensional
  hyperentanglement through a biphoton frequency comb}.
\newblock \bibinfo{journal}{Nature Photonics} \emph{\bibinfo{volume}{9}},
  \bibinfo{pages}{536--542}.
\bibitem[{Chapman et~al.(2020)Chapman, Graham, Zeitler, Bernstein and
  Kwiat}]{chapman2020time}
\bibinfo{author}{Chapman, J.C.}, \bibinfo{author}{Graham, T.M.},
  \bibinfo{author}{Zeitler, C.K.}, \bibinfo{author}{Bernstein, H.J.}, and
  \bibinfo{author}{Kwiat, P.G.} (\bibinfo{year}{2020}).
  \bibinfo{title}{Time-bin and polarization superdense teleportation for space
  applications}.
\newblock \bibinfo{journal}{Physical Review Applied}
  \emph{\bibinfo{volume}{14}}, \bibinfo{pages}{014044}.
\bibitem[{Jiang et~al.(2025)Jiang, Yan, Lu, Chen, Wen, An, Chen, Lu, Zhu and
  Ma}]{jiang2025entanglement}
\bibinfo{author}{Jiang, Z.}, \bibinfo{author}{Yan, W.}, \bibinfo{author}{Lu,
  C.}, \bibinfo{author}{Chen, Y.}, \bibinfo{author}{Wen, W.},
  \bibinfo{author}{An, Y.Y.}, \bibinfo{author}{Chen, L.}, \bibinfo{author}{Lu,
  Y.}, \bibinfo{author}{Zhu, S.}, and \bibinfo{author}{Ma, X.S.}
  (\bibinfo{year}{2025}). \bibinfo{title}{Entanglement distribution over
  metropolitan fiber using on-chip broadband polarization entangled photon
  source}.
\newblock \bibinfo{journal}{arXiv preprint arXiv:2503.07198}.
\bibitem[{Yin et~al.(2019)Yin, Serafini, Su, Shiue, Timurdogan, Fanto and
  Preble}]{yin2019low}
\bibinfo{author}{Yin, P.}, \bibinfo{author}{Serafini, J.R.},
  \bibinfo{author}{Su, Z.}, \bibinfo{author}{Shiue, R.J.},
  \bibinfo{author}{Timurdogan, E.}, \bibinfo{author}{Fanto, M.L.}, and
  \bibinfo{author}{Preble, S.} (\bibinfo{year}{2019}). \bibinfo{title}{Low
  connector-to-connector loss through silicon photonic chips using ultra-low
  loss splicing of smf-28 to high numerical aperture fibers}.
\newblock \bibinfo{journal}{Optics express} \emph{\bibinfo{volume}{27}},
  \bibinfo{pages}{24188--24193}.
\bibitem[{Harada et~al.(2009)Harada, Takesue, Fukuda, Tsuchizawa, Watanabe,
  Yamada, Tokura and Itabashi}]{harada2009frequency}
\bibinfo{author}{Harada, K.i.}, \bibinfo{author}{Takesue, H.},
  \bibinfo{author}{Fukuda, H.}, \bibinfo{author}{Tsuchizawa, T.},
  \bibinfo{author}{Watanabe, T.}, \bibinfo{author}{Yamada, K.},
  \bibinfo{author}{Tokura, Y.}, and \bibinfo{author}{Itabashi, S.i.}
  (\bibinfo{year}{2009}). \bibinfo{title}{Frequency and polarization
  characteristics of correlated photon-pair generation using a silicon wire
  waveguide}.
\newblock \bibinfo{journal}{IEEE Journal of Selected Topics in Quantum
  Electronics} \emph{\bibinfo{volume}{16}}, \bibinfo{pages}{325--331}.
\bibitem[{Ma et~al.(2016)Ma, Sacher, Tang, Mikkelsen, Yang, Xu, Thiessen, Lo
  and Poon}]{ma2016silicon}
\bibinfo{author}{Ma, C.}, \bibinfo{author}{Sacher, W.D.},
  \bibinfo{author}{Tang, Z.}, \bibinfo{author}{Mikkelsen, J.C.},
  \bibinfo{author}{Yang, Y.}, \bibinfo{author}{Xu, F.},
  \bibinfo{author}{Thiessen, T.}, \bibinfo{author}{Lo, H.K.}, and
  \bibinfo{author}{Poon, J.K.} (\bibinfo{year}{2016}). \bibinfo{title}{Silicon
  photonic transmitter for polarization-encoded quantum key distribution}.
\newblock \bibinfo{journal}{Optica} \emph{\bibinfo{volume}{3}},
  \bibinfo{pages}{1274--1278}.
\bibitem[{Zhang et~al.(2024{\natexlab{b}})Zhang, Wu and
  Poon}]{zhang2024polarization}
\bibinfo{author}{Zhang, Q.}, \bibinfo{author}{Wu, K.}, and
  \bibinfo{author}{Poon, A.W.} (\bibinfo{year}{2024}{\natexlab{b}}).
  \bibinfo{title}{Polarization entanglement generation in silicon nitride
  waveguide-coupled dual microring resonators}.
\newblock \bibinfo{journal}{Optics Express} \emph{\bibinfo{volume}{32}},
  \bibinfo{pages}{22804--22816}.

\end{thebibliography}

\end{document}